\documentclass{article}
\usepackage{amssymb, amsmath}
\numberwithin{equation}{section}
\newtheorem{theorem}{Theorem}[section]
\newtheorem{proposition}{Proposition}[section]
\newtheorem{lemma}{Lemma}[section]
\newtheorem{corollary}{Corollary}[section]
\newtheorem{remark}{Remark}[section]
\newtheorem{definition}{Definition}[section]
\begin{document}
\title{Global Existence of Solutions for the Einstein-Boltzmann
system in a Bianchi Type I Space-Time(Detailed paper)}
\author{Norbert NOUTCHEGUEME$^{1}$ and David DONGO$^{1}$ \\
$^{1}$Department of Mathematics, Faculty of Sciences, University of Yaounde
1,\\
PO Box 812, Yaounde, Cameroon \\
{e-mail: nnoutch@justice.com, nnoutch@uycdc.uninet.cm }}
\date{}
\maketitle

\begin{abstract}
We prove a global in time existence theorem, for the initial value
problem for the Einstein-Boltzmann system, with arbitrarily large
initial data, in the homogeneous case, in a Bianchi Type I
space-time.
\end{abstract}

\section{Introduction}

The Einstein equations are the basic equations of the General
Relativity, theory funded in 1916 by Albert Einstein and according
to which: ''the geometric structure of the space-time is
determined by its material and energetic contents''. These
equations link, on one hand, the geometry of the space-time, which
is determined by a fundamental tool: the metric tensor, standing
for the gravitational field, and, on the other hand, all the
material and energetic contents, summarized by the
stress-energy-matter tensor (we will call it the matter tensor),
acting as the sources for the gravitational field. Solving the
Einstein equations is determining both the gravitational field and
its sources. Naturally, the problem varies with the matter
contents, and this is why, in the mathematical study of General
Relativity, one of the main problems is to solve the Einstein
equations coupled to various field equations, in order to
determine both the gravitational field and its sources in
different cases. A fundamental problem today is to establish the
existence and to give the properties of global solutions to such
coupled systems. We are interested in global dynamics of
relativistic kinetic matter, and our goal is to prove such results
in that domain. In the case of \textbf{collisionless} matter, the
governing system is the Einstein-Vlasov system, in the pure
gravitational case, or this system coupled to other field
equations, if these fields are
involved in the sources of the Einstein equations. In the \textbf{%
collisionless} case, several authors proved global results, see \cite{a},
\cite{b} for reviews, \cite{c}, \cite{d} and \cite{e} for scalar matter
fields, also see \cite{g}, \cite{h} for the Einstein-Vlasov system with
cosmological constant, which turns out to be a useful tool for the proof of
the fact that, the expansion of the universe is accelerating, see \cite{q}
for more details on this question.\newline
Now, in the case of \textbf{collisional matter}, the Einstein-Vlasov system
is replaced by the Einstein-Boltzmann system, which seems to be the best
approximation available and that describes the case of instantaneous, binary
and elastic collisions. In contrast with the abundance of papers in the
collisionless case, the literature seems very poor in the \textbf{collisional%
} case. If, due to its importance in collisional kinetic theory, several
authors proved global results for the single Boltzmann equation, see \cite{i}%
, \cite{j}, \cite{k} for the non-relativistic case, and \cite{l}
for the full relativistic case, very few authors studied the
Einstein-Boltzmann system, see \cite{m} for a local existence
theorem. It then seems interesting to us, to extend to the
\textbf{collisional} case some global results obtained in the
\textbf{collisionless} case. This was certainly the objective of
the author in \cite{n}, \cite{o}, in which he studied the
existence of global solutions for the Einstein-Boltzmann system.
Unfortunately, several points of the work are far from clear, such
as, the use of a formulation which is valid only for the
non-relativistic Boltzmann equation, or, concerning the Einstein
equations, to abandon the evolution equations, which are really
relevant in the spatially homogeneous case he considers, to
concentrate on the constraints equations, that reduce, as we will
see, to a simple question of choice of the initial data. In
\cite{w}, the authors prove a recent result on the global
existence of solutions for the spatially homogeneous
Einstein-Boltzmann system on a Robertson-Walker space-time, in the
case of a strictly positive Cosmological constant $\Lambda >0$.

In this papers, we study the collisional evolution of a kind of
\textbf{uncharged} massive particles, under the only influence of
their common gravitational field, represented by the metric tensor
we denote by $g$, and
which is a function of the position of the particles, and whose components $%
g_{\alpha\beta}$, sometimes called \newline ''gravitational
potentials'', are subject to the Einstein equations. Here, the
sources of the Einstein equations are generated by the only matter
contents which are the massive particles, through the matter
tensor which is a symmetric 2-tensor, we denote $T_{\alpha\beta}$.
The particles are statistically described in terms of their
distribution function, we denote by $f$, and which is a
non-negative real-valued function of both the position and the
momentum of particles. The scalar function $f$ which is physically
interpreted as the ''probability of the presence density'' of the
massive particles, during their collisional evolution, is subject
to the Boltzmann equation, defined by a non-linear operator called
the '' collision operator''. In the binary and elastic scheme due
to Lichnerowicz and Chernikov (1940), we adopt, at a given
position, only 2 particles collide each other, in an instantaneous
shock affecting only the momentum of each particle, only the sum
of the 2 momenta being preserved. We then study the coupled
Einstein-Boltzmann system in $(g,f)$. The system is coupled in the
sense that each unknown appears in the equation determining the
other unknown: $f$ which is subject to the Boltzmann equation
generates the
sources of the Einstein equations through the matter tensor $T_{\alpha\beta}$%
, whereas the metric tensor $g$ which is subject to the Einstein equations
is in both sides of the Boltzmann equation in $f$. If the particles were
\textbf{charged}, the above system would be coupled to the Maxwell equations
that would describe the electromagnetic effects; this is our project in a
future work.

We now specify the geometric framework, i.e. the kind of
space-time we are looking for in the present work. An important
part of the General Relativity is Cosmology, which is the study of
the universe at large scale; in such a view-point, even cluster of
galaxies are assimilable to ''particles''. A. Einstein and W. de
Sitter introduced cosmological models in 1917.\newline Let us
point out that the Einstein equations are overdetermined and
physically meaning symmetriy assumptions reduce the number of
unknown. The surface symmetry assumptions which are the spherical,
plane and hyperbolic symmetries assumptions constitute the major
part of the models studied in the works we quote in the
references. Robertson and Walker showed in 1944 that '' Exact
spherical symmetry about every point would imply that the universe
is spatially homogeneous'' see \cite{p}, p. 135. In the
Robertson-Walker models, the spatial geometry has constant
curvature which is positive, zero and negative respectively. We
look for a spatially homogeneous, locally rotationally symmetric
(LRS) Bianchi type I space-time, which is a direct generalization
of the Robertson-Walker space-time with zero curvature and which
is known, in Cosmology, to be the basic model for the study of the
expanding universe. In the model we are looking for, the metric
tensor $g$ has only two unknown components $a$ and $b$, and the
spatial homogeneity means that $a$ and $b$ depend only on the time
$t$, and the distribution function $f$ depends only on the time
$t$ and the momentum $p$ of the particles. The study of the
Einstein-Boltzmann system then turns out to the determination of
the triplet of scalar functions $(a,b,f)$, $a$ and $b $ being
strictly positive functions.

We now sketch the strategy we adopt to prove the global in time
existence of a solution $(a,b,f)$ of the initial value problem for
the coupled
Einstein-Boltzmann system, for arbitrarily large initial data $%
(a_{0},b_{0},f_{0})$ at the initial time $t=0$. In the spatially homogeneous
case we consider, the Einstein equations are a system of three non-linear
o.d.e in $a$ and $b$. The Boltzmann equation is a non-linear first order
p.d.e for the distribution function $f$.

In the first step, we suppose $a$ and $b$ given with the only assumption to
be bounded away from zero; and we give, following Glassey R.T. in \cite{r},
the correct formulation of the relativistic Boltzmann equation in $f$ on a
Bianchi type I space-time. We then prove that, on any bounded time interval %
\mbox{$I=[t_{0}, t_{0}+T]$}, with $t_{0}\in \mathbb{R}_{+}$, $T \in %
\mathbb{R}_{+}^{*}$, the initial value problem for the Boltzmann
equation, with initial data $f_{t_{0}}\in
L_{2}^{1}(\mathbb{R}^{3})$, $f_{t_{0}}\geq 0 $ a.e; has a unique
solution \mbox{$f \in C[I; L_{2}^{1}(\mathbb{R}^{3})]$}, $f(t)\geq
0$ a.e, $\forall t \in I$, where $L_{2}^{1}(\mathbb{R}^{3})$ is a
weighted subspace of $L^{1}(\mathbb{R}^{3})$, whose weight , is
imposed by the manner in which the sources terms $T_{\alpha\beta}$
of the Einstein equations depend on $f$. We follow the method
developed in \cite{s} to prove
the global existence of the solution $f \in C[0, +\infty;\, L^{1}(\mathbb{R}%
^{3})]$ for the initial value problem for the Boltzmann equation,
but here the weighted norm of $L_{2}^{1}(\mathbb{R}^{3})$ could
allow us to prove the existence theorem only on bounded time
intervals $I=[t_{0}, t_{0}+T]$, but this was enough for the
coupling with the Einstein equations and to obtain the global
existence by another method.

In a second step, we suppose $f$ given in $C[I;
L_{2}^{1}(\mathbb{R}^{3})]$ and we consider the Einstein equations
in $a$ and $b$, that split into the constraint equations and the
evolution equations. The constraint equations contain the momentum
constraints which are automatically satisfied in the homogeneous
case we consider, and the Hamiltonian constraint which reduces to
a question of choice for the initial data. The main problem is
then to
solve the evolution equations which are a system of two non-linear o.d.e in $%
a$ and $b$. We set, following Rendall, A.D., and Uggla, C. in \cite{x}
\[
H = -\frac{tr\,k}{3}; \quad z = \frac{1}{a^{-2}+2b^{-2}+1} ; \quad s = \frac{%
b^{2}}{b^{2}+2a^{2}} ; \quad \Sigma _{+} = -\frac{3}{tr\,k}\frac{\stackrel{.%
}{b}}{b}-1.
\]
in which $k$ is the second fundamental form on the space-like
3-manifolds that foliate the space-time in the 3+1 formulation of
the Einstein equations. We then prove that the Einstein evolution
system in $a$ and $b$ is equivalent to a first order system in
$(H, z, s, \Sigma _{+})$, for which standard theorems apply. From
there, we deduce the existence of the solutions $a,\,b$ of the
initial value problem for the Einstein evolution equations, on any
interval $I=[0, T[$, $T>0$, which are increasing function of $t$,
and hence bounded from below, satisfying consequently the
assumptions of the first step.

In a third step, we use the change of variables indicated above together
with the characteristic method for the first order p.d.e that reduces the
Boltzmann equation to a non linear o.d.e in $f$ with values in the Banach
space $L_{2}^{1}(\mathbb{R}^{3})$, to prove the \textbf{local} existence of
solutions for the coupled Einstein-Boltzmann system by applying standard
results.

In a fourth step, we use the results of the two first steps to construct an
appropriate functional framework, allowing us to prove, using the fixed
point theorem that, the local solution of the coupled system, whose
existence is established in the third step, is in fact, a \textbf{global}
solution.

Our method preserves the physical nature of the problem that imposes to the
distribution function $f$ to be a non-negative function, and this is why we
had to consider spaces of functions $f$ which are positive everywhere.
Nowhere we had to require smallness assumptions on the initial data, which
can, consequently, be taken arbitrarily large.

An aspect of the results of the present paper is an extension of the results
of \cite{w}, to the case of the zero cosmological constant $\Lambda=0$.
Notice that the present paper follows papers \cite{s} and \cite{w} and the
proofs for the Boltzmann equation are similar, but profound differences
arise with \cite{w} on the Einstein equations.

This paper is organized as follows:\newline
In section 2, we introduce the Einstein equations, the Boltzmann equation
and the coupled Einstein-Boltzmann system, on a Bianchi type I space-time.%
\newline
In section 3, we study the Boltzmann equation in $f$.\newline
In section 4, we study the Einstein equations in $a$ and $b$.\newline
In section 5, we prove the existence of local solutions to the coupled
Einstein-Boltzmann system.\newline
In section 6, we prove the existence of global solutions, to the coupled
Einstein-Boltzmann system.

\section{The Boltzmann equation, the Einstein equations and the
Einstein-Boltzmann system on a LRS Bianchi type I space-time.}

\subsection{Notations and function spaces}

A greek index varies from 0 to 3 and a Latin index from 1 to 3,
unless otherwise specified. We adopt the Einstein summation
convention $a_{\alpha }b^{\alpha } \equiv \underset{\alpha
}{\sum}a_{\alpha }b^{\alpha }$. We consider a locally rotationally
symmetric (LRS) Bianchi type I space-time denoted
($\mathbb{R}^{4}$, $g$) where, for $x=(x^{\alpha
})=(x^{0},x^{i})\in \mathbb{R}^{4}$, $x^{0}=t$ denotes the time and $\bar{x}%
=(x^{i})$ the space; $g$ stands for the metric tensor with
signature \mbox{(-, +, +, +)} that can be written:
\begin{equation}\label{eq:2.1}
g=-dt^{2}+a^{2}(t)(dx^{1})^{2}+b^{2}(t)[(dx^{2})^{2}+(dx^{3})^{2}]
\end{equation}
in which $a$ and $b$ are strictly positive functions of $t$. We
consider the
\textbf{collisional} evolution of a kind of \textbf{uncharged} \textbf{%
massive} relativistic particles in the time oriented space-time $(\mathbb{R}%
^{4},g)$. The particles are statistically described by their \textbf{%
distribution} \textbf{function} we denote $f$, which is a
non-negative real-valued function of both the position $(x^{\alpha
})$ and the 4-momentum $p=(p^{\alpha })$ of the particles, and
that coordinatize the tangent bundle $T(\mathbb{R}^{4})$ i.e:
\[
f:T(\mathbb{R}^{4})\simeq \mathbb{R}^{4}\times \mathbb{R}^{4}\rightarrow %
\mathbb{R}_{+},\quad (x^{\alpha },p^{\alpha })\mapsto f(x^{\alpha
},p^{\alpha })\in \mathbb{R}_{+}
\]
For $\bar{p}$ = $(p^{i})$, $\bar{q}$ = $(q^{i})$ $\in
\mathbb{R}^{3}$, we
set, as usual $|\bar{p}|=[\underset{i}{\sum}(p^{i})^{2}]^{%
\frac{1}{2}}$,  and we define the scalar product:
\begin{equation}\label{eq:2.2}
<\bar{p}.\bar{q}>=a^{2}p^{1}q^{1}+b^{2}[p^{2}q^{2}+p^{3}q^{3}]
\end{equation}
We suppose the rest mass $m>0$ of the particles normalized to
unity, i.e $m=1$. The relativistic particles are then required to
move on the future sheet of the mass-shell, whose equation is
$g(p,p)=-1$; from which we deduce, using expression (\ref{eq:2.1})
of $g$:
\begin{equation}\label{eq:2.3}
p^{0}=\sqrt{1+a^{2}(p^{1})^{2}+b^{2}[(p^{2})^{2}+(p^{3})^{2}]}
\end{equation}
where the choice of $p^{0}>0$ symbolizes the fact that, in a
natural manner, particles eject towards the future. (\ref{eq:2.3})
shows that, in fact, $f$ is defined on the 7-dimensional subbundle
of $T(\mathbb{R}^{4})$ coordinatized by $(x^{\alpha }),(p^{i})$.
In the  spatially homogeneous case we consider, $f$ depends only
on $t$ and $\bar{p}=(p^{i})$.\newline The framework we will refer
to is the subspace of $L^{1}(\mathbb{R}^{3})$, denote
$L_{2}^{1}(\mathbb{R}^{3})$ and defined by :
\begin{equation}  \label{eq:2.4}
L_{2}^{1}(\mathbb{R}^{3})=\{f\in L^{1}(\mathbb{R}^{3}),\Vert f\Vert :=\int_{%
\mathbb{R}^{3}}\sqrt{1+|\bar{p}|^{2}}|f(\bar{p})|d\bar{p}<+\infty \}
\end{equation}
$\Vert .\Vert $ is a norm on $L_{2}^{1}(\mathbb{R}^{3})$ and $(L_{2}^{1}(%
\mathbb{R}^{3}),\Vert .\Vert )$ is a Banach space. We set for $r$ an
arbitrary strictly positive real number;
\begin{equation}\label{eq:2.5}
X_{r}=\{f\in L_{2}^{1}(\mathbb{R}^{3}),\quad f\geq 0\quad a.e,\quad \Vert
f\Vert \leq r\}
\end{equation}
One verifies easily that, endowed with the metric induced by the norm $\Vert
.\Vert $, $X_{r}$ is a complete and connected metric subspace of $(L_{2}^{1}(%
\mathbb{R}^{3}),\Vert .\Vert )$. Let $I$ be an real interval; we set:
\[
C[I;\,L_{2}^{1}(\mathbb{R}^{3})]=\{f:I\rightarrow L_{2}^{1}(\mathbb{R}^{3}),%
\text{ f continuous and bounded }\}
\]
endowed with the norm:
\begin{equation} \label{eq:2.6}
\|| f \|| = \underset{t \in I}{Sup}\|f(t)\|
\end{equation}
$C[I;\,L_{2}^{1}(\mathbb{R}^{3})]$ is a Banach space. $X_{r}$
being defined by (\ref{eq:2.5}); we set:
\begin{equation}\label{eq:2.7}
C[I;X_{r}]=\{f\in C[I;\,L_{2}^{1}(\mathbb{R}^{3})],\,f(t)\in
X_{r},\quad \forall t\in I\}
\end{equation}
Endowed with the metric induced by the norm $\Vert |.\Vert |$ defined by (%
\ref{eq:2.6}), $C[I;X_{r}]$ is a complete metric subspace of
$\left( C[I;L_{2}^{1}(\mathbb{R}^{3})],\,|\Vert .|\Vert \right) $.

\subsection{The Boltzmann equation}

In its general form, the Boltzmann equation in $f$, on the curved space-time %
\mbox{($\mathbb{R}^{4},\, g$)} can be written:
\begin{equation}  \label{eq:2.8}
p^{\alpha}\frac{\partial{f}}{\partial{x^{\alpha}}} -
\Gamma^{\alpha}_{\mu\nu}p^{\mu}p^{\nu}\frac{\partial{f}}{\partial{p^{\alpha}}%
} = Q(f, f)
\end{equation}
Where ($\Gamma_{\lambda\mu}^{\alpha})$ are the Christoffel symbols
of $g$, and Q is a non-linear integral operator called the
''Collision Operator''. We specify this operator in details in
next section. Now, since $f$ depends only on $t$ and $(p^{i})$,
(\ref{eq:2.8}) writes :
\begin{equation}  \label{eq:2.9}
p^{0}\frac{\partial{f}}{\partial{t}} - \Gamma^{i}_{\mu\nu}p^{\mu}p^{\nu}%
\frac{\partial{f}}{\partial{p^{i}}} = Q(f, f)
\end{equation}
We now express the Christoffel symbols in the case of the metric $g$ defined
by (\ref{eq:2.1}). Their general expression is:
\begin{equation}  \label{eq:2.10}
\Gamma^{\lambda}_{\alpha\beta}= \frac{1}{2}g^{\lambda\mu}[\partial_{%
\alpha}g_{\mu\beta} + \partial_{\beta}g_{\alpha\mu} -
\partial_{\mu}g_{\alpha\beta}]
\end{equation}
in which ($g^{\lambda\mu}$) denotes the inverse matrix of ($g_{\lambda\mu}$%
); We have $\Gamma^{\lambda}_{\alpha\beta}=\Gamma^{\lambda}_{\beta\alpha}$.
The definition (\ref{eq:2.1}) of $g$ gives at once:
\begin{equation}  \label{eq:2.11}
\left\{
\begin{array}{ll}
g^{00} = g_{00} = -1; \quad g_{11} = a^{2};\quad g_{22} = g_{33} = b^{2};
\quad g^{11} = a^{-2}; \quad g^{22} = g^{33} = b^{-2} &  \\
g_{0i} = g^{0i} = 0; \quad g_{ij} = g^{ij} = 0 \quad for \quad i \neq j &
\end{array}
\right.
\end{equation}
A direct computation, using (\ref{eq:2.10}) and (\ref{eq:2.11}), gives, with
$\dot{a} = \frac{da}{dt}$:
\begin{equation}  \label{eq:2.12}
\left\{
\begin{array}{ll}
\Gamma_{11}^{0} = \dot{a}a; \quad\Gamma_{22}^{0} = \Gamma_{33}^{0} = \dot{b}%
b; \quad \Gamma_{10}^{1} = \frac{\dot{a}}{a}; \quad \Gamma_{20}^{2} =
\Gamma_{30}^{3} = \frac{\dot{b}}{b}; \quad \Gamma_{00}^{0} = 0; &  \\
\Gamma_{\alpha\beta}^{0} = 0 \quad for \quad \alpha \neq \beta; \quad
\Gamma_{ij}^{k} = 0 &
\end{array}
\right.
\end{equation}
The Boltzmann equation (\ref{eq:2.10}) then writes, using (\ref{eq:2.12}):
\begin{equation}  \label{eq:2.13}
\frac{\partial{f}}{\partial{t}} - 2\frac{\dot{a}}{a}p^{1}\frac{\partial{f}}{%
\partial{p^{1}}} - 2\frac{\dot{b}}{b}p^{2}\frac{\partial{f}}{\partial{p^{2}}}
- 2\frac{\dot{b}}{b}p^{3}\frac{\partial{f}}{\partial{p^{3}}} = \frac{1}{p^{0}%
} Q(f, f)
\end{equation}
in which $p^{0}$ is given by (\ref{eq:2.3}). (\ref{eq:2.13}) is a non-linear
p.d.e in $f$ we study in next section.

\subsection{The Einstein Equations}

The Einstein equations in $g = (g_{\alpha\beta})$ can be written:
\begin{equation}  \label{eq:2.14}
R_{\alpha\beta} - \frac{1}{2} R\, g_{\alpha\beta} = 8 \pi G T_{\alpha\beta}
\end{equation}
in which:\newline $R_{\alpha\beta}$ is the Ricci tensor ,
contracted of the curvature tensor of $g$ we specify
bellow;\newline $R = g^{\alpha\beta}R_{\alpha\beta} =
R^{\alpha}_{\alpha}$ is the scalar curvature;\newline
$T_{\alpha\beta}$ is the matter tensor that represents the matter
contents, namely the massive particles in our case.
$T_{\alpha\beta}$ is generated by the distribution function $f$ of
the particles by:
\begin{equation}  \label{eq:2.15}
T^{\alpha\beta}(t) = \int_{\mathbb{R}^{3}}\frac{p^{\alpha}p^{\beta}f(t, \bar{%
p})|g|^{\frac{1}{2}}}{p^{0}}d\bar{p}
\end{equation}
where $|g|$ is the determinant of $g$; (\ref{eq:2.1}) gives $|g|^{\frac{1}{2}%
} = ab^{2}$. Recall that $f$ is a function of $t$ and $\bar{p} = (p^{i})$;
hence $T^{\alpha\beta}$ is a function of $t$.\newline
$G$ is the universal gravitational constant. We take $G = 1$.

The contraction of the Bianchi identities gives the identities $%
\nabla_{\alpha}S^{\alpha\beta} = 0$, where $S_{\alpha\beta} =
R_{\alpha\beta} - \frac{1}{2} R\, g_{\alpha\beta}$ is the Einstein tensor.
The Einstein equations (\ref{eq:2.14}) then imply that the tensor $%
T^{\alpha\beta}$ must satisfy the four relations $\nabla_{\alpha}T^{\alpha%
\beta} = 0$ called the conservation laws. But it is proved in
\cite{t} that these laws are satisfied for all solution $f$ of the
Boltzmann equation. We have:
\begin{equation}  \label{eq:2.16}
\left\{
\begin{array}{ll}
R_{\alpha\beta} = R_{\,\,\,\alpha, \lambda\beta}^{\lambda} &  \\
where &  \\
R_{\mu\,\,\,,\alpha\beta}^{\,\,\,\lambda} = \partial_{\alpha}{\Gamma^{\lambda}_{\mu%
\beta}} - \partial_{\beta}{\Gamma^{\lambda}_{\mu\alpha}} +
\Gamma^{\lambda}_{\nu\alpha}\Gamma^{\nu}_{\mu\beta}-
\Gamma^{\lambda}_{\nu\beta}\Gamma^{\nu}_{\mu\alpha} &
\end{array}
\right.
\end{equation}
in which $\Gamma^{\lambda}_{\mu\beta}$ is given by (\ref{eq:2.10}) with the
only non zero components given by (\ref{eq:2.12}). (\ref{eq:2.12}) and (\ref
{eq:2.16}) show that the Einstein equations (\ref{eq:2.14}) are a system of
second order non-linear o.d.e in $a$ and $b$. In order to have things fresh
in mind at the adequate moment, we leave the expression of (\ref{eq:2.14})
in term of $a$ and $b$ to section 4, which is devoted to the study of the
Einstein equations.

\subsection{The coupled Einstein-Boltzmann system}

(\ref{eq:2.14})-(\ref{eq:2.13}) is the Einstein-Boltzmann system
we study. The system is coupled in the sense that, $f$ which is
subject to the Boltzmann equation (\ref{eq:2.13}) generates
through (\ref{eq:2.15}), the sources terms $T_{\alpha\beta}$ of
the Einstein equations; it appears
clearly that the coefficients of the Boltzmann equation (\ref{eq:2.13}) in $%
f $ depend on $a$ and $b$ which are subject to the Einstein equations (\ref
{eq:2.14}), and we will see, in next section, that $a$ and $b$ also appear
in the collision operator $Q$, which is the r.h.s of the Boltzmann equation (%
\ref{eq:2.13}).

\section{Existence Theorem for the Boltzmann Equation}

In this section, we suppose that the components $a$ and $b$ of the metric
tensor $g$ are given and fixed, and we prove an existence theorem for the
initial values problem for the Boltzmann equation (\ref{eq:2.13}), on every
bounded interval $I = [t_{0}, t_{0} + T]$ where $t_{0} \in \mathbb{R}_{+}$, $%
T \in \mathbb{R}_{+}^{*}$. We begin by specifying the collision operator $Q$
in (\ref{eq:2.13})

\subsection{The Collision Operator}

In the instantaneous, binary and elastic scheme, due to Lichnerowicz and
Chernikov we consider, at a given position $(t,\bar{x})$, only 2 particles
collide instantaneously, without destroying each other, the collision
affecting only the momentum of each particle that changes after the shock,
only the sum of the 2 momenta being preserved, following the scheme:

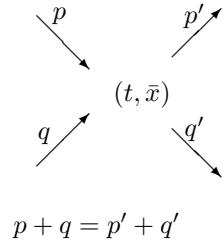
\newcounter{cms} \setlength{\unitlength}{1mm}

\begin{center}
\begin{picture}(10,20)
\put(11,6.5){\makebox(0,0){$(t,\bar{x})$}}
\put(-3,-3){\vector(1,1){7}} \put(-3,17){\vector(1,-1){7}}
\put(15,2){\vector(1,-1){6.5}} \put(15,11.5){\vector(1,1){6.5}}
\put(-2,1){\makebox(0,0){$q$}} \put(0,17){\makebox(0,0){$p$}}
\put(18,2){\makebox(0,0){$q'$}} \put(18,17){\makebox(0,0){$p'$}}
\end{picture}
\end{center}

\[
p + q = p^{\prime}+ q^{\prime}
\]
The collision operator $Q$ is defined, using functions $f$, $g$ on $%
\mathbb{R}^{3}$, $p,q$ standing for the momenta \textbf{before}
the collision and $p^{\prime},q^{\prime}$ the momenta
\textbf{after} the collision by:
\begin{equation}  \label{eq:3.1}
Q(f, g) = Q^{+}(f, g) - Q^{-}(f, g)
\end{equation}
where:
\begin{equation}  \label{eq:3.2}
\quad Q^{+}(f, g)(\bar{p}) = \int_{\mathbb{R}^{3}} \frac{ab^{2}d\bar{q}}{%
q^{0}} \int_{S^{2}}f(\bar{p^{\prime}})g(\bar{q^{\prime}} )A(a, b, \bar{p},
\bar{q}, \bar{p^{\prime}}, \bar{q^{\prime}}) d\omega
\end{equation}
\begin{equation}  \label{eq:3.3}
\quad Q^{-}(f,g)(\bar{p}) =\int_{\mathbb{R}^{3}}. \frac{ab^{2}d\bar{q}}{q^{0}%
}\int_{S^{2}}f(\bar{p})g(\bar{q})A(a, b, \bar{p}, \bar{q}, \bar{p^{\prime}},
\bar{q^{\prime}})d\omega
\end{equation}
whose elements we now introduce step by step, specifying
properties and hypotheses we adopt:

\begin{itemize}
\item[1)]  $S^{2}$ is the unit sphere of $\mathbb{R}^{3}$, whose volume
element is denoted $dw$.

\item[2)]  A is a non-negative real-valued regular function of all
its arguments, called the \textbf{collision kernel} or the
\textbf{cross-section} of the collisions, on which we require the
following boundedness, symmetry and Lipschitz continuity
assumptions:
\begin{equation}  \label{eq:3.4}
0\leq A(a,\,b,\, \bar{p},\, \bar{q},\, \bar{p^{\prime}},\, \bar{q^{\prime}})
\leq C_{0}
\end{equation}
\begin{equation}  \label{eq:3.5}
A(a,\,b,\, \bar{p},\, \bar{q},\, \bar{p^{\prime}},\, \bar{q^{\prime}}) =
A(a,\,b,\, \bar{p^{\prime}},\, \bar{q^{\prime}},\, \bar{p},\, \bar{q})
\end{equation}
\begin{equation}  \label{eq:3.6}
A(a,\,b,\, \bar{p},\, \bar{q},\, \bar{p^{\prime}},\, \bar{q^{\prime}}) =
A(a,\,b,\, \bar{q},\, \bar{p},\, \bar{q^{\prime}},\, \bar{p^{\prime}})
\end{equation}
\begin{equation}  \label{eq:3.7}
|A(a_{1},\,b,\, \bar{p},\, \bar{q},\, \bar{p^{\prime}},\, \bar{q^{\prime}})
- A(a_{2},\,b,\, \bar{p},\, \bar{q},\, \bar{p^{\prime}},\, \bar{q^{\prime}})
| \leq k_{0}|a_{1} - a_{2}|
\end{equation}
\begin{equation}  \label{eq:3.8}
|A(a,\,b_{1},\, \bar{p},\, \bar{q},\, \bar{p^{\prime}},\, \bar{q^{\prime}})
- A(a,\,b_{2},\, \bar{p},\, \bar{q},\, \bar{p^{\prime}},\, \bar{q^{\prime}})
| \leq k_{0}|b_{1} - b_{2}|
\end{equation}
where $C_{0}$ and $k_{0}$ are strictly positive constants.

\item[3)]  The conservation law $p + q = p^{\prime}+ q^{\prime}$ splits
into:
\begin{equation}\label{eq:3.9}
p^{0} + q^{0} = p^{'0} + q^{'0}
\end{equation}
\begin{equation}  \label{eq:3.10}
\bar{p} + \bar{q} = \bar{p^{\prime}} + \bar{q^{\prime}}
\end{equation}
and (\ref{eq:3.9}) shows, using (\ref{eq:2.3}), the conservation of the
quantity:
\begin{equation}  \label{eq:3.11}
e = \sqrt{1 + a^{2}(p^{1})^{2} + b^{2}[(p^{2})^{2}+(p^{3})^{2}]}+ \sqrt{1 +
a^{2}(q^{1})^{2} + b^{2}[(q^{2})^{2}+(q^{3})^{2}]}
\end{equation}
called the elementary energy of the unit rest mass particles; we can
interpret (\ref{eq:3.10}) by setting, following Glassey, R. T., in \cite{r}:
\begin{equation}  \label{eq:3.12}
\begin{cases} \bar{p'} = \bar{p} + R(\bar{p}, \bar{q}, \omega )\omega\\
\bar{q'} = \bar{q} - R(\bar{p}, \bar{q}, \omega)\omega \, ; \qquad \omega
\in S^{2} \end{cases}
\end{equation}
in which $R(\bar{p}, \bar{q}, \omega )$ is a real-valued function. We prove,
by a straightforward computation (using (\ref{eq:2.3}) to express ${%
p^{\prime}}^{0}$, ${q^{\prime}}^{0}$ in terms of $\bar{p^{\prime}}$, $\bar{%
q^{\prime}}$, and next (\ref{eq:3.12}) to express $\bar{p^{\prime}}$, $\bar{%
p^{\prime}}$ in terms of $\bar{p}$, $\bar{p}$, $\omega$, $R$, that equation (%
\ref{eq:3.9}) leads to a quadratic equation in R, that solves to give the
only non trivial solution:
\begin{equation}  \label{eq:3.13}
R(\bar{p}, \bar{q}, \omega) = \frac{2\,p^{o}q^{o} e\,< \omega , (\hat{\bar{q}%
} - \hat{\bar{p}})>}{e^{2} - [<\omega.(\bar{p} + \bar{q})]^{2}}
\end{equation}
in which $\hat{\bar{p}} = \frac{\bar{p}}{p^{0}}$, $e$ is given by (\ref{eq:3.11}%
), and $<,>$ is the scalar product defined by (\ref{eq:2.2}). Another direct
computation gives, using the properties of the determinants, that the
Jacobian of the change of variables $(\bar{p}, \bar{q}) \rightarrow (\bar{%
p^{\prime}}, \bar{q^{\prime}})$ in $\mathbb{R}^{3}\times\mathbb{R}^{3}$,
defined by (\ref{eq:3.12}) is:
\begin{equation}  \label{eq:3.14}
\frac{\partial(\bar{p'}, \bar{q'})}{\partial(\bar{p}, \bar{q}%
)} = -\frac{p'^{o}q'^{o}}{p^{o}q^{o}}
\end{equation}
\end{itemize}
(\ref{eq:3.14}) shows, using once more (\ref{eq:2.3}) and the implicit
functions theorem, that the change of variables (\ref{eq:3.12}) is
invertible and also allows to compute $\bar{p}$, $\bar{q}$ in terms of $\bar{%
p^{\prime}}$, $\bar{q^{\prime}}$. \newline
Finally, formulae (\ref{eq:2.3}) and (\ref{eq:3.12}) show that the functions
to integrate in (\ref{eq:3.2}) and (\ref{eq:3.3}) completely express in
terms of $\bar{p}, \bar{q}, \omega$, and the integrations with respect to $%
\bar{q}$ and $\omega$ leave functions $Q^{+}(f, g)$, $Q^{-}(f, g)$ in terms
of the single variable $\bar{p}$. In practice, we will consider functions $%
f: \mathbb{R} \times \mathbb{R}^{3}\rightarrow \mathbb{R}$, that induce, for
$t$ fixed in $\mathbb{R}$, functions $f(t)$ on $\mathbb{R}^{3}$, defined by $%
f(t)(\bar{p}) = f(t, \bar{p})$. Even in such cases, in order to simplify
notations, we will explicit the dependence on $t$ only if necessary.

\begin{remark}
\begin{itemize}
\item[1)]  Formulae (\ref{eq:3.13}) and (\ref{eq:3.14}) are generalizations
to the case of the LRS Bianchi type I space-time, of analogous formulae
established by Glassey, R.T, in \cite{r}, in the case of the Minkowski
space-time, which corresponds to the case $a(t) = b(t)= 1$ in the metric
defined by (\ref{eq:2.1}), in which case the scalar product $<,>$ defined by
(\ref{eq:2.2}) reduces to the usual scalar product in $\mathbb{R}^{3}$.

\item[2)]  In (\ref{eq:3.2}), (\ref{eq:3.3}),
$\omega_{q}=|g|^{\frac{1}{2}}
\frac{d\bar{q}}{q^{0}}=ab^{2}\frac{d\bar{q}}{q^{0}}$ is a Leray
form, which is in fact, the canonical volume element in the
momenta space.

\item[3)]  $A = k e^{- a^{2}-b^{2} - |\bar{p}|^{2} - |\bar{q}|^{2} - |\bar{%
p^{\prime}}|^{2} - |\bar{q^{\prime}}|^{2}}$,\, $k > 0$, is a simple example
of collision kernel satisfying assumptions (\ref{eq:3.4}), (\ref{eq:3.5}), (%
\ref{eq:3.6}), (\ref{eq:3.7}) and (\ref{eq:3.8}).
\end{itemize}
\end{remark}

\subsection{Resolution of the Boltzmann equation}

We consider the Boltzmann equation (2.13) on $I=[t_{0}, t_{0} +
T]$, $t_{0} \in \mathbb{R}_{+}$, $T >0$.

The functions $a$ and $b$ are supposed to be defined on $I$.
Equation (\ref{eq:2.13}) is a first order p.d.e in $f$ and its
resolution is equivalent to the resolution of the associated
characteristic system, which can be written, taking t as
parameter:
\begin{equation}  \label{eq:3.15}
\frac{dp^{1}}{dt} = -2\,\frac{\dot{a}}{a}p^{1}; \quad \frac{dp^{2}}{dt} =
-2\,\frac{\dot{b}}{b}p^{2}; \quad \frac{dp^{3}}{dt} = -2\,\frac{\dot{b}}{b}%
p^{3}; \quad \frac{df}{dt} = \frac{1}{p^{0}}Q(f, f)
\end{equation}
We solve the initial value problem on $I = [t_{0}, t_{0} + T]$,
with initial data:
\begin{equation}  \label{eq:3.16}
p^{i}(t_{0}) = y^{i}; i=1, 2, 3; \qquad f(t_{0}) = f_{t_{0}}
\end{equation}
The equations in $\bar{p} = (p^{i})$ solve directly to give,
setting $y = (y^{i}) \in \mathbb{R}^{3}$;
\begin{equation}  \label{eq:3.17}
\bar{p}(t_{0} + t, y) =D(t)y, \,\, \text{with} \,\, D(t)= Diag\left(\frac{%
a^{2}(t_{0})}{a^{2}(t_{0} + t)}, \frac{b^{2}(t_{0})}{b^{2}(t_{0} +
t)}, \frac{b^{2}(t_{0})}{b^{2}(t_{0} + t)}\right), \,\, t\in [0,
T]
\end{equation}
which shows that, the initial value problem for the Boltzmann
equation (2.13) is finally equivalent to the following integral
equation in which, for simplicity, we conserve the notation
$\bar{p}$, which stands this time for any independent variable in
$\mathbb{R}^{3}$:
\begin{equation}  \label{eq:3.17}
f(t_{0} + t, \bar{p}) = f_{t_{0}}(\bar{p}) + \int_{t_{0}}^{t_{0} + t}\frac{1%
}{p^{0}}Q(f, f)(s, \bar{p})ds \quad t \in [0, T]
\end{equation}
We prove:
\begin{theorem}
\label{t:3.2} Let $a$ and $b$ be  strictly positive continuous
functions of $t$, such that $a(t) \geq \frac{3}{2}$, $b(t)\geq
\frac{3}{2}$, whenever $a$
and $b$ are defined . Let $f_{t_{0}} \in L_{2}^{1}(\mathbb{R}^{3})$, $%
f_{t_{0}} \geq 0$, a.e, and $r \in \mathbb{R}^{*}_{+}$ such that
$r > \|f_{t_{0}}\|$.\newline Then, the initial value problem for
the Boltzmznn equation on $[t_{0}, t_{0} + T]$, with initial data
$f_{t_{0}}$, has a unique solution $f \in C[t_{0}, t_{0} + T;
X_{r}]$. Moreover, f satisfies the estimation:
\begin{equation}  \label{eq:3.18}
\underset{t \in [t_{0}, t_{0} + T]}{Sup}\|f(t)\| \leq
\|f_{t_{0}}\|
\end{equation}
\end{theorem}
\textbf{Proof:} Theorem \ref{t:3.2} is a direct consequence of the
following result:
\begin{proposition}
\label{p:3.3} Assume hypotheses of theorem \ref{t:3.2} on: $a$, $b$, $%
f_{t_{0}}$ and $r$.

\begin{itemize}
\item[1)]  There exists an integer $n_{0}(r)\geq1$ such that, for every
integer $n \geq n_{0}(r)$ and for every $v \in X_{r}$, the equation:
\begin{equation}  \label{eq:3.19}
\sqrt{n}\, u - \frac{1}{p^{0}}Q(u, u) = v
\end{equation}
has a unique solution $u_{n} \in X_{r}$.

\item[2)]  Let $n \in \mathbb{N}$, $n \geq n_{0}(r)$

\begin{itemize}
\item[i)]  For every $u \in X_{r}$ , define R(n, Q)u to be the unique
element of $X_{r}$ such that:
\begin{equation}  \label{eq:3.20}
\sqrt{n}R(n, Q)u - \frac{1}{p^{0}}Q\left[R(n, Q)u, R(n, Q)u\right] = u
\end{equation}

\item[ii)]  Define operator $Q_{n}$ on $X_{r}$ by:
\begin{equation}  \label{eq:3.21}
Q_{n}(u, u) = n\sqrt{n}R(n, Q)u - n u
\end{equation}
Then

\begin{itemize}
\item[a)]  The integral equation
\begin{equation}  \label{eq:3.22}
f(t_{0} + t, \bar{p}) = f_{t_{0}}(\bar{p}) + \int_{t_{0}}^{t_{0} +
t}Q_{n}(f, f)(s, \bar{p})ds \quad t\in[0, T]
\end{equation}
has a unique solution $f_{n} \in C[t_{0}, t_{0} + T; X_{r}]$. Moreover, $%
f_{n}$ satisfies the estimation:
\begin{equation}  \label{eq:3.23}
\underset{t \in [t_{0}, t_{0} + T]}{Sup}\|f_{n}(t)\| \leq
\|f_{t_{0}}\|
\end{equation}

\item[b)]  The sequence $(f_{n})$ converges in $C[t_{0}, t_{0} + T; X_{r}]$%
to an element $f \in C[t_{0}, t_{0} + T; X_{r}]$, which is the unique
solution of the integral equation (\ref{eq:3.17}). The solution $f$
satisfies the estimation (\ref{eq:3.18}).
\end{itemize}
\end{itemize}
\end{itemize}
\end{proposition}

\textbf{Proof of the proposition \ref{p:3.3}:}\newline
As we pointed it out, the above result is similar to those of \cite{s} and
\cite{w}. The proof of prop \ref{p:3.3} above is analogous to the proof of
prop.3.1 of \cite{w}, and it follows the same lines as the proof of theorem
4.1 in \cite{s}. We will emphasize only on points where differences arise
with the present case, and, at the same time, we establish useful formulae
for next steps of the present paper.

\textbf{A) Proof of point 1) of prop 3.1}\newline
We use:

\begin{lemma}
\label{l:3.4} Let $f, g \in L^{1}_{2}(\mathbb{R}^{3})$. then $\frac{1}{p^{0}}%
Q^{+}(f, g), \, \, \frac{1}{p^{0}}Q^{-}(f, g) \in L^{1}_{2}(\mathbb{R}^{3})$
and
\begin{equation}  \label{eq:3.24}
\left \| \frac{1}{p^{0}}Q^{+}(f, g) \right \| \leq C(t)\parallel f \parallel
\parallel g \parallel, \quad \left \| \frac{1}{p^{0}}Q^{-}(f, g) \right \|
\leq C(t)\parallel f \parallel \parallel g \parallel
\end{equation}

\begin{equation}  \label{eq:3.25}
\begin{cases} \left \| \frac{1}{p^{0}}Q^{+}(f, f) - \frac{1}{p^{0}}Q^{+}(g,
g)\right \| \leq C(t)(\parallel f \parallel + \parallel g
\parallel)\parallel f - g \parallel\\ \left \| \frac{1}{p^{0}}Q^{-}(f, f) -
\frac{1}{p^{0}}Q^{-}(g, g)\right \| \leq C(t)(\parallel f \parallel +
\parallel g \parallel)\parallel f - g \parallel \end{cases}
\end{equation}
\begin{equation}  \label{eq:3.26}
\left \| \frac{1}{p^{0}}Q(f, f) - \frac{1}{p^{0}}Q(g, g)\right \| \leq
C(t)(\parallel f \parallel + \parallel\ g \parallel) \parallel f - g
\parallel
\end{equation}
where
\begin{equation}  \label{eq:3.27}
C(t)= C ab^{2}(t)
\end{equation}
in which $C > 0$ is an absolute constant.
\end{lemma}

\textbf{proof of lemma 3.1}\newline
We deduce from (\ref{eq:3.9}), using $a > 1$, $b > 1$ and expression (\ref
{eq:2.3}) of $p^{0}$ :
\[
\sqrt{1 + |\bar{p}|^{2}} \leq \sqrt{1 +
\frac{1}{a^{2}}+\frac{1}{b^{2}}} p^{0} \leq 2(p^{0} + q^{0}) =
2(p'^{0} + q'^{0})
\]
Expression (\ref{eq:3.2}) of $Q^{+}(f, g)$ then gives, since by (\ref{eq:3.4}%
), $|A| \leq C_{0}$:
\[
\begin{aligned} \left\|\frac{1}{p^{0}}Q^{+}(f, g)\right\| &=
\int_{\mathbb{R}^{3}}\left|\frac{\sqrt{1 + |\bar{p}|^{2}}}{p^{0}} Q^{+}(f,
g)\right|d\bar{p}\\ &\leq
2ab^{2}(t)C_{0}\int_{\mathbb{R}^{3}}\int_{\mathbb{R}^{3}}\int_{S^{2}}%
\frac{p'^{0} +
q'^{0}}{p^{0}q^{0}}|f(\bar{p'})||g(\bar{q'})|d\bar{p}d\bar{q}d\omega
\end{aligned}
\]
We make the change of variable (\ref{eq:3.12}) and  we deduce from
(\ref{eq:3.14})
that gives $d\bar{p}d\bar{q} \, = \, \frac{%
p^{0}q^{0}}{p'^{0}q'^{0}} \,d\bar{p'}d\bar{q' }$, and since by
(\ref{eq:2.3}) we have $\frac{p'^{0} + q'^{0}}{p'^{0}q'^{0}} =
\frac{1}{q'^{0}} + \frac{1}{q'^{0}} \leq 2$, that the above
inequality yields:
\[
\left\|\frac{1}{p^{0}}Q^{+}(f, g)\right\| \leq 8\pi ab^{2}(t)C_{0}\int_{%
\mathbb{R}^{3}}\int_{\mathbb{R}^{3}} |f(\bar{p^{\prime}})||g(\bar{q^{\prime}}%
)|d\bar{p^{\prime}}d\bar{q^{\prime}} \leq C ab^{2}\| f \|\| g\|
\]
with $C=8\pi C_{0}$. The estimation of
$\left\|\frac{1}{p^{0}}Q^{-}(f, g)\right\|$ follows the same way
without change of variables and (\ref{eq:3.24}) follows. The
inequalities (\ref{eq:3.25}) are direct consequences of
(\ref{eq:3.24}) and the bilinearity of $Q^{+}$ and $Q^{-}$, which
allows us to write, $P$ standing for $\frac{1}{p^{0}}Q^{+}$ or
$\frac{1}{p^{0}}Q^{-}$:
\[
P(f, f) - P(g, g) = P(f, f- g) + P(f - g, g).
\]
Finally, (\ref{eq:3.26}) is a consequence of (\ref{eq:3.25}) and $Q = Q^{+}
- Q^{-}$. This completes the proof of the lemma 3.1$\blacksquare$

Now, since the function $t \rightarrow ab^{2}(t)$ is positive and
continuous on
the line segment $[t_{0}, t_{0} + T]$, there exists an absolute constant $%
C(t_{0}, T) > 0$ such that $C(t)\leq C(t_{0}, T) $, $\forall t \in
[t_{0}, t_{0} + T]$, where $C(t)$ is defined by (\ref{eq:3.27}).
We then obtain, from lemma 3.1, the same inequalities with an
absolute constant, than the inequalities in proposition 3.1 in
\cite{s} . The proof of the point 1) of proposition 3.1 above is
then exactly the same as the proof of proposition 3.2 in \cite{s}.
$\blacksquare$\newline

\textbf{B) Proof of the point 2) of prop 3.1}\newline
We use, $n_{0}(r)$ being the integer defined in point 1) of proposition 3.1
above :

\begin{lemma}
\label{l:3.5} We have, for every integer $n \geq n_{0}(r)$ and for every $u
\in X_{r}$
\begin{equation}  \label{eq:3.28}
\parallel \sqrt{n}R(n, Q)u\parallel = \parallel u \parallel
\end{equation}
\end{lemma}

\textbf{Proof of the lemma 3.2}\newline
(\ref{eq:3.28}) is a consequence of :
\begin{equation}  \label{eq:3.29}
\int_{\mathbb{R}^{3}}Q(f, f)(\bar{p})d \bar{p} = 0, \quad \forall
f\in L^{1}_{2}(\mathbb{R}^{3})
\end{equation}
(\ref{eq:3.29}) is established exactly as formula (3.30) in the
proof of lemma 3.5 in \cite{w}, replacing everywhere in that
proof, $a^{3}$ by $ab^{2}$, and since we did, in the present
paper, the same assumptions and we have the same properties as
those used in \cite{w}. Now deduce (\ref {eq:3.28}) from
(\ref{eq:3.29}) exactly as in the proof of lemma 3.5 in \cite {w},
choosing this time $B = Diag (a, b, b)$ instead of $B = Diag (a,
a, a)$ and using $a \geq \frac{3}{2}$, $b \geq \frac{3}{2}$, to
conclude the proof of lemma 3.2 above.

Now (\ref{eq:3.28}) is exactly the equality (\ref{eq:3.10}) in proposition
3.3 in \cite{s}. We then prove exactly as for proposition 3.3 in \cite{s},
that all the other relations hold in the present case. The proofs of points
2) a) and 2) b) of the above \mbox{proposition 3.1} are exactly the same as
the proofs of \mbox{proposition
  4.1} and theorem 4.1 in \cite{s}, just replacing $[0, + \infty[$ by $%
[t_{0}, t_{0} + T]$. This completes the proof of \mbox{proposition 3.1}
above, which gives directly theorem 3.1 $\blacksquare$

\section{Existence Theorem of the Einstein Equations}

\subsection{Expression and Reduction of the Einstein Equations}

We express the Einstein equations (\ref{eq:2.14}) in terms of $a$ and $b$.
We have to compute the Ricci tensor $R_{\alpha\beta}$ given by (\ref{eq:2.16}%
). The expression (2.12) of $\Gamma^{\lambda}_{\alpha\beta}$ shows, using
(2.16), that the only non-zero components of the Ricci tensor are the $%
R_{\alpha\alpha}$ and  $R_{22} = R_{33}$. Then, it will be enough
to compute $R_{00} = R^{\lambda}_{\,\,0,\lambda 0 }$; $R_{11} =
R^{\lambda}_{\,\,1, \lambda 1}$ and $R_{22} = R^{\lambda}_{\,\,2, \lambda 2}$%
. The expression (\ref{eq:2.12}) of $\Gamma_{\alpha\beta}^{\lambda}$ and
formulae (\ref{eq:2.16}) give:
\[
R^{0}_{\,\,0, 0 0} = 0; \quad R^{1}_{\,\,0, 1 0} =-\frac{\ddot{a}}{a};\quad
R^{2}_{\,\,0, 2 0} = R^{3}_{\,\,0,3 0} = - \frac{\ddot{b}}{b};
\]
\[
R^{0}_{\,\,1, 0 1} = a\ddot{a}; \quad R^{1}_{\,\,1,1 1} = 0; \quad
R^{2}_{\,\,1,2 1} = R^{3}_{\,\,1,3 1} = a\dot{a}\frac{\dot{b}}{b}
\]
\[
R^{0}_{\,\,2, 0 2} = b\ddot{b}; \quad R^{1}_{\,\,2,1 2} = b\dot{b}\frac{\dot{%
a}}{a}; \quad R^{2}_{\,\,2,2 2} =0; \quad R^{3}_{\,\,2, 3 2} = (\dot{b})^{2}
\]
We then have:
\[
R_{00} = - \frac{\ddot{a}}{a}- 2\frac{\ddot{b}}{b}; \quad R_{11} = a\ddot{a}
+ 2a\dot{a}\frac{\dot{b}}{b}; \quad R_{22} = R_{33} = b\ddot{b} + b\dot{b}%
\frac{\dot{a}}{a}+(\dot{b})^{2}
\]
We can then compute the scalar curvature $R$ to be:
\[
R = g^{\alpha\beta}R_{\alpha\beta} = 2\left[\frac{\ddot{a}}{a} + 2\frac{%
\ddot{b}}{b} + 2\frac{\dot{a}\dot{b}}{ab} + \left( \frac{\dot{b}}{b}
\right)^{2}\right]
\]
The Einstein equations (\ref{eq:2.14}) in $a$ and $b$ then take the reduced
form:
\begin{equation}  \label{eq:4.1}
2\frac{\dot{a}\dot{b}}{ab} + \left( \frac{\dot{b}}{b} \right)^{2} = 8 \pi
T_{00}
\end{equation}
\begin{equation}  \label{eq:4.2}
-a^{2}\left[ 2\frac{\ddot{b}}{b} + \left( \frac{\dot{b}}{b} \right)^{2}
\right] = 8 \pi T_{11}
\end{equation}
\begin{equation}  \label{eq:4.3}
-b^{2}\left[\frac{\ddot{a}}{a} + \frac{\ddot{b}}{b} + \frac{\dot{a}\dot{b}}{%
ab} \right] = 8 \pi T_{22}
\end{equation}
in which $T_{\alpha\beta} = g_{\alpha\lambda}g_{\beta\mu}T^{\lambda\mu}$ is
given in terms of $f$ by (\ref{eq:2.15}).\newline
In this paragraph, we suppose $f$ fixed in $C[I; X_{r}]$, where $I = [0, T]$%
, $T > 0$, with $f(0) = f_{0} \in L_{2}^{1}(\mathbb{R}^{3})$,
$f_{0} \geq 0$ a.e. and $r > \|f_{0}\|$, $r$ fixed. We study the
initial value problem for the non-linear second order system
(\ref{eq:4.1})-(\ref{eq:4.2})- (\ref {eq:4.3}), in $a$ and $b$,
with initial data $a_{0}, \,b_{0}, \,\dot{a}_{0}, \, \dot{b}_{0};
$ i.e
\begin{equation}  \label{eq:4.4}
a(0) = a_{0}; \quad b(0) = b_{0}; \quad \dot{a}(0)=\dot{a}_{0}; \quad \dot{b}%
(0)=\dot{b}_{0}
\end{equation}

\subsection{Compatibility}

We notice that, if $S_{\alpha\beta} = R_{\alpha\beta} - \frac{1}{2}%
g_{\alpha\beta}R$ is the Einstein tensor, we have
\[
S_{0i} = 0; \quad S_{ij} = 0 \quad for \quad i\neq j; \quad S_{22} = S_{33}
\]
then, the compatibility of the Einstein equations (\ref{eq:2.14}) require
that:
\begin{equation}  \label{eq:4.5}
T_{0i} = 0; \quad T_{ij} = 0 \quad for \quad i \neq j; \quad T_{22} = T_{33}
\end{equation}
But the matter tensor $T_{\alpha\beta} =
g_{\alpha\lambda}g_{\beta\mu}T^{\lambda\mu}$ is defined by (\ref{eq:2.15})
in terms of the distribution function $f$. It then appears that (\ref{eq:4.5}%
) are in fact conditions to impose to $f$. we prove:

\begin{proposition}
\label{p:4.1} Let $G$ be the sub-group of $O(3)$ whose elements are on the
form:
\[
M_{\epsilon}(\theta) = \left(
\begin{array}{ccc}
\epsilon & 0 & 0 \\
0 & \cos \theta & -\sin \theta \\
0 & \sin \theta & \cos \theta
\end{array}
\right); \quad \epsilon^{2} = 1, \quad \theta \in \mathbb{R}
\]
Assume the hypotheses of theorem 3.1 for $t_{0}=0$. Assume in addition that $%
f_{0}$ is invariant by $G$, and that the collision kernel $A$ satisfies:
\begin{equation}  \label{eq:4.6}
A(a, b, M\bar{p} , M\bar{q}, M\bar{p^{\prime}} ,
M\bar{q^{\prime}}) = A(a, b , \bar{p} , \bar{q} , \bar{p^{\prime}}
, \bar{q^{\prime}}), \quad \forall \bar{p},\bar{q}\in
\mathbb{R}^{3}, \quad \forall M \in G
\end{equation}
then
\begin{itemize}
\item[1)]  The solution $f$ of the integral equation (\ref{eq:3.17}) on $[0,
T]$ satisfies:
\begin{equation}  \label{eq:4.7}
f( t, M\bar{p}) = f( t, \bar{p}) \quad \forall t \in [0, T], \quad
\forall \bar{p} \in \mathbb{R}^{3}, \qquad \forall M \in G
\end{equation}

\item[2)]  The matter tensor $T_{\alpha\beta}$ satisfies the conditions (\ref
{eq:4.5}).
\end{itemize}
\end{proposition}

\textbf{Proof of Proposition 4.1:} Let $M \in G$\newline
1) One verifies easily, that the scalar product $<,>$ defined by (\ref
{eq:2.2}) and, consequently, $p^{0}(\bar{p})$ defined by (\ref{eq:2.3}) are
invariant by $G$, which means \newline
$<M\bar{p}, M\bar{q}> = <\bar{p}, \bar{q}>$, $p^{0}(M\bar{p}) = p^{0}(\bar{p}%
)$, $\forall M \in G$. Now (\ref{eq:3.17}) gives, since $f_{0}(M\bar{p}) =
f_{0}(\bar{p})$:
\begin{equation}  \label{eq:4.8}
f(t , M\bar{p}) = f_{0}(\bar{p}) + \int_{0}^{ t}\frac{1}{p^{0}}Q(f, f)(s, M%
\bar{p})ds ,\quad \forall t \in [0, T]
\end{equation}
Next, definition (\ref{eq:3.1}), (\ref{eq:3.2}), (\ref{eq:3.3}) of the
collision operator $Q$ gives:
\begin{equation}  \label{eq:4.9}
Q(f, f)(s, M\bar{p}) = \int_{\mathbb{R}^{3}}\frac{ab^{2}}{q^{0}}d\bar{q}
\int_{S^{2}}[f(s, \bar{p^{\prime}})f(s, \bar{q^{\prime}}) - f(s, M\bar{p}%
)f(s,\bar{q}) ]A(a , b, M\bar{p} , \bar{q} , \bar{p^{\prime}} , \bar{%
q^{\prime}})d\omega
\end{equation}
Let us set in (\ref{eq:4.9}) $\bar{q} = M \bar{q}_{1}$; $\omega = M
\omega_{1}$. Then, formulae (\ref{eq:3.12}) give, using expression (\ref
{eq:3.13}) of $R$ and the invariance of the scalar product $<,>$ by the
subgroup $G$ of $O(3)$, $\forall M \in G$:
\[
\left\{
\begin{array}{ll}
\bar{p^{\prime}} = M\bar{p} + R(M\bar{p}, \bar{q}, \omega)\omega = M\bar{p}
+ R(M\bar{p}, M\bar{q}_{1}, M\omega_{1})M\omega_{1} = M\left(\bar{p} + R(%
\bar{p}, \bar{q}_{1}, \omega_{1})\omega_{1}\right) &  \\
\bar{q^{\prime}} = \bar{q} - R(M\bar{p}, \bar{q}, \omega)\omega = M\bar{q}%
_{1} - R(M\bar{p}, M\bar{q}_{1}, M\omega_{1})M\omega_{1} = M\left(\bar{q}%
_{1} - R(\bar{p}, \bar{q}_{1}, \omega_{1})\omega_{1}\right) &
\end{array}
\right.
\]
So that $\bar{p^{\prime}} = M \bar{p^{\prime}}_{1}$; $\bar{q^{\prime}} = M
\bar{q^{\prime}}_{1}$ where:
\[
\bar{p^{\prime}}_{1} = \bar{p} + R(\bar{p}, \bar{q}_{1},
\omega_{1})\omega_{1}; \quad \bar{q^{\prime}}_{1} = \bar{q}_{1}-R(\bar{p}, \bar{q%
}_{1}, \omega_{1})\omega_{1}.
\]
Then (\ref{eq:4.9}) implies, using assumption (\ref{eq:4.6}) on $A$, $q^{0}
= q_{1}^{0}$, and the invariance of the volume elements $d\bar{q}$, $d\omega$
by $G$:
\[
Q(f, f)(s,\,M\bar{p}) = Q[f(s)oM, f(s)oM](\bar{p})
\]
(\ref{eq:4.8}) then writes:
\[
\left(f(t )oM\right)(\bar{p}) = f_{0}(\bar{p}) + \int_{0}^{ t}\frac{1}{p^{0}}%
Q[f(s)oM, f(s)oM](\bar{p})ds
\]
which shows, by setting $h(s) = f(s)oM$, that $\|h(s)\| =
\|f(s)\|$ (by setting $\bar{q} = M\bar{p}$ in the integral
defining $\|f(s)oM\|$) , and that, $h$ and $f$ are 2 solutions in
$C[0, T; X_{r}]$ of the integral equation (\ref{eq:3.17}). The
uniqueness theorem 3.1 then implies that $h = f $ and the  point 1
of proposition 4.1 is proved.\newline 2) Consider the expression
(\ref{eq:2.15}) of $T^{\alpha\beta}$ in which $f$ satisfies
(\ref{eq:4.7}) and observe that, if $\bar{p} = M\bar{q}$ where $M
\in G$, then $p^{0}(\bar{p}) = p^{0}(M\bar{q}) = p^{0}(\bar{q}) =
q^{0}$

\begin{itemize}
\item[(i)]  Set in (\ref{eq:2.15}) $\alpha = 0$, $\beta = i$ with i = 1, 2,
3; now compute the integral using the change of variables $\bar{p} =
M_{-1}(\pi)\bar{q}$; where $M_{\epsilon}(\theta)$ is defined in prop.4.1;
the integral in $\bar{q}$ gives $T^{0i} = - T^{0i}$, i = 1, 2, 3; then $%
T^{0i} = 0 $, i = 1, 2, 3 and this implies $T_{0i} = 0$, i = 1, 2, 3.

\item[(ii)]  Set in (\ref{eq:2.15}) $\alpha = i$, $\beta = j$, $i \neq j$;
now compute the integral using the change of variables $\bar{p} = M_{-1}(0)%
\bar{q}$ if ($i = 1, j = 2$) or ($i = 1, j = 3$) and $\bar{p} = M_{1}(\frac{%
\pi}{2})\bar{q}$ if ($i = 2, j = 3$), to obtain: $T^{ij} = - T^{ij}$ if $%
i\neq j$; then $T^{12} = T^{13} =T^{23} = 0 $ and $T_{12} = T_{13} = T_{23}
= 0$.

\item[(iii)]  Set in (\ref{eq:2.15}) $\alpha = \beta = 2$ and compute the
integral using the change of variables $\bar{p} = M_{1}(\frac{\pi}{2})\bar{q}
$. The integral in $\bar{q}$ gives $T^{22} = T^{33}$ and hence $T_{22} =
T_{33}$.
\end{itemize}

This completes the proof of proposition 4.1 $\blacksquare$

In all what follows, we assume that the collision kernel $A$
satisfies assumption (\ref{eq:4.6}), and that $f(t)$ is invariant
by $G$ . When we will study the coupled Einstein-Boltzmann system,
we know by prop.4.1 that it will be enough that $f_{0}$ be
invariant by $G$. Also notice that the kernel $A$ defined in
remark 3.1 is an example of kernel satisfying assumption
(\ref{eq:4.6}).

\subsection{The Constraint Equations}

The Einstein equations (\ref{eq:2.14}) in which $G = 1$, give, using $%
S^{\alpha\beta}$:
\begin{equation}  \label{eq:4.10}
S^{0}_{\alpha} - 8 \pi T^{0}_{\alpha} = 0
\end{equation}
It is proved in \cite{u}, p.39, that, in the general case, the
four quantities:\\ \mbox{$H^{0}_{\alpha} = S^{0}_{\alpha} - 8 \pi
 T^{0}_{\alpha}$} do not involve the second derivative with respect to $t$,
of the metric tensor $g_{\alpha\beta}$ and, using the identities $%
\nabla_{\alpha}(S^{\alpha\beta} - 8\pi T^{\alpha\beta}) = 0$, that the
quantities $H^{0}_{\alpha}$ satisfy a linear homogeneous first order
differential system. Consequently:

\begin{itemize}
\item[1${{}^\circ}$)]  For $t = 0$, the quantities $S^{0}_{\alpha} - 8 \pi
T^{0}_{\alpha}$ express uniquely in terms of the initial data $a_{0}, \dot{a}%
_{0}, b_{0}, \dot{b}_{0}$ and $f_{0}$.

\item[2${{}^\circ}$)]  If $(S^{0}_{\alpha} - 8 \pi T^{0}_{\alpha})(0) = 0$,
then $(S^{0}_{\alpha} - 8 \pi T^{0}_{\alpha})(t) = 0$ in the whole existence
domain of the solution $(a, b)$ of the Einstein equations (\ref{eq:4.1})-(%
\ref{eq:4.2})-(\ref{eq:4.3}). In other words, (\ref{eq:4.10}) is
satisfied if the initial data satisfied \textbf{the constraint}
\[
S^{0}_{\alpha}(0) = 8\pi T^{0}_{\alpha}(0)
\]
\end{itemize}

For this reason, the equations (\ref{eq:4.10}) are called constraint
equations. Notice that, since $S^{0}_{i} = 0$ and $T^{0}_{i} = 0$ by
prop.4.1, equations (\ref{eq:4.10}) for $\alpha = i$, which are called
\textbf{momentum constraints}, are automatically satisfied. It then remains equation (%
\ref{eq:4.10}) for $\alpha = 0$ which is called the
\textbf{Hamiltonian constraint}. Now $S^{0}_{0} - 8\pi T^{0}_{0} =
0$ is equivalent to:
\begin{equation}  \label{eq:4.11}
S^{00} = 8\pi T^{00}
\end{equation}
which is exactly (\ref{eq:4.1}), since $S_{00} = S^{00}$; $T_{00} =T^{00}$.
So, (\ref{eq:4.1}) is the Hamiltonian constraint which is satisfied if it is
the case for $t = 0$, i.e, if the initial data $a_{0}, b_{0}, \dot{a}_{0},
\dot{b}_{0}, f_{0}$ satisfy, using (\ref{eq:2.15}) for $\alpha = \beta = 0$,
(\ref{eq:4.1}), (\ref{eq:4.11}), the initial constraint
\begin{equation}  \label{eq:4.12}
2\frac{\dot{a}_{0}\dot{b}_{0}}{a_{0}b_{0}} + \left( \frac{\dot{b}_{0}}{b_{0}}
\right)^{2} = 8\pi\int_{\mathbb{R}^{3}}p^{0}(0)f_{0}(\bar{p})a_{0}b_{0}^{2}d%
\bar{p}
\end{equation}
with, using (\ref{eq:2.3}) and (\ref{eq:4.4}), $p^{0}(0) = \sqrt{1 +
a_{0}^{2}(p^{1})^{2} + b_{0}^{2}[(p^{2})^{2}+(p^{3})^{2}]}$. It appears
that, the choice of the four independent initial data $a_{0}, b_{0}, \dot{b}%
_{0}, f_{0}$ uniquely determines $\dot{a}_{0}$. We will take:
\begin{equation}  \label{eq:4.13}
a_{0}>0,\,\, b_{0}>0,\,\, \dot{b}_{0}>0,\,\, f_{0}\in L^{1}_{2}(\mathbb{R}%
^{3}),\,\,f_{0}\geq 0\,\, \text{a.e};\, \, \dot{a}_{0}>0.
\end{equation}
where $\dot{a}_{0}>0$ is obtained by taking $b_{0}$ sufficiently large.%
\newline
Notice that, in the 3+1 formulation of the Einstein equations, the
Hamiltonian constraint (\ref{eq:4.1}) writes:
\begin{equation}  \label{eq:4.14}
(tr K)^{2} - K_{ij}K^{ij} = 16\pi T_{00}
\end{equation}
where $K = (K_{ij})$ is the second fundamental form induced by the
metric $g$ on the hypersurfaces $S_{t} = \{t\}\times
\mathbb{R}^{3}$. $K_{ij}$ is given in the present case by $K_{ij}
= -\frac{1}{2}\partial_{t}g_{ij}$, which gives, using
(\ref{eq:2.1})
\begin{equation}  \label{eq:4.15}
K_{11} = -a\dot{a}; \quad K_{22} = K_{33} = -b\dot{b}; \quad
K_{ij} = 0 \quad if \quad i \neq j
\end{equation}
and $tr K = g^{ij}K_{ij}$ is the trace of $K$; $tr K$ which
represents the mean curvature of the space-time is then given by:
\begin{equation}  \label{eq:4.16}
tr K = -\left( \frac{\dot{a}}{a}+ 2\frac{\dot{b}}{b} \right)
\end{equation}

\subsection{The Evolution Equations}

The Einstein evolution equations are the second order non-linear
o.d.e (\ref {eq:4.2})-(\ref{eq:4.3}) in $a$ and $b$. we study the
initial values problem
for this system on $[0, T]$, $T>0$ with initial data defined by (\ref{eq:4.4}%
) and satisfying the initial constraint (\ref{eq:4.12}). As we explained
above, we are funded to consider the Hamiltonian constraint (\ref{eq:4.1})
or (\ref{eq:4.14}), as auxilliary equation.

In order to apply standard results, we are going to show that the Einstein
evolution equations are equivalent to a non-linear first order system. We
consider the system (\ref{eq:4.1})-(\ref{eq:4.2})-(\ref{eq:4.3}) and we set
in the sources terms, with $T_{\alpha\beta}$ defined by (\ref{eq:2.15}):
\begin{equation}  \label{eq:4.17}
\rho = 8\pi T_{00}
\end{equation}
\begin{equation}  \label{eq:4.18}
P_{1} = \frac{8\pi T_{11}}{a^{2}}
\end{equation}
\begin{equation}  \label{eq:4.19}
P_{2} = \frac{8\pi T_{22}}{b^{2}}
\end{equation}
\begin{equation}  \label{eq:4.20}
R = \frac{P_{1}+2P_{2}}{\rho}
\end{equation}
\begin{equation}  \label{eq:4.21}
R_{+} = \frac{P_{2}-P_{1}}{\rho}
\end{equation}
Next we consider, following Rendall,A.D and Uggla,C. in \cite{x}, the change
of variables:
\begin{equation}  \label{eq:4.22}
H = -\frac{tr K}{3}
\end{equation}
\begin{equation}  \label{eq:4.23}
z = \frac{1}{a^{-2}+2b^{-2}+1}
\end{equation}
\begin{equation}  \label{eq:4.24}
s = \frac{b^{2}}{b^{2}+2a^{2}}
\end{equation}
\begin{equation}  \label{eq:4.25}
\Sigma _{+} = \frac{1}{H}\frac{\dot{b}}{b} - 1
\end{equation}
$H$ is called the Hubble variable. Finally we set:
\begin{equation}  \label{eq:4.26}
\Omega = \frac{\rho}{3H^{2}}
\end{equation}
\begin{equation}  \label{eq:4.27}
q = 2\Sigma_{+}^{2} + \frac{\Omega}{2}(1+R)
\end{equation}
$\Omega$ is called the normalized energy density and $q$, the deceleration
parameter.

We have the following immediate consequences of the above definitions

\begin{lemma}
\label{l:4.2}
\begin{equation}  \label{eq:4.28}
0<z<1; \, 0<s<1; \, a^{2}=\frac{z}{s(1-z)}; \, b^{2}=\frac{2z}{(1-s)(1-z)};
\, \Omega \geq 0;\, \Omega = 1-\Sigma_{+}^{2}
\end{equation}
\begin{equation}  \label{eq:4.29}
0\leq P_{1}+2P_{2}\leq \rho; \quad 0\leq R \leq 1; \quad 0\leq q \leq 2
\end{equation}
\begin{equation}  \label{eq:4.30}
\left\{
\begin{array}{ll}
p^{0}(s,z) =\sqrt{1+\frac{z}{s(1-z)}(p^{1})^{2}+ \frac{2z}{(1-s)(1-z)} [%
(p^{2})^{2}+(p^{3})^{2}]} &  \\
P_{1} = \frac{16\pi z^{\frac{5}{2}}}{s^{\frac{3}{2}}(1-s)(1-z)^{\frac{5}{2}}}
\int_{\mathbb{R}^{3}}\frac{(p^{1})^{2}\, f}{p^{0}(s,z)}d\bar{p} &  \\
P_{2} = \frac{32\pi z^{\frac{5}{2}}}{s^{\frac{1}{2}}(1-s)^{2}(1-z)^{\frac{5}{%
2}}} \int_{\mathbb{R}^{3}}\frac{(p^{2})^{2}\, f}{p^{0}(s,z)}d\bar{p} &
\end{array}
\right.
\end{equation}
\end{lemma}

\textbf{Proof}

\begin{itemize}
\item  i) (\ref{eq:4.23})-(\ref{eq:4.24}) imply $0<z<1$, $0<s<1$ and solving
this system in $a^{2}$, $b^{2}$, yields the solutions given in (\ref{eq:4.28}%
). Next, $(f(t)\geq 0, a.e)$ implies \mbox{$\rho = 8 \pi
T_{00}\geq 0$}; (\ref {eq:4.26}) then shows that $\Omega \geq 0$.
For the last result in (\ref
{eq:4.28}), consider (\ref{eq:4.14}) that writes, using (\ref{eq:4.17}) and (%
\ref{eq:4.26}):\\ \mbox{$(tr K)^{2}-K_{ij}K^{ij}=6\Omega H^{2}$},
in which, by (\ref {eq:4.15})
\begin{equation}  \label{eq:4.31}
K_{ij}K^{ij}=\left( \frac{\dot{a}}{a} \right)^{2}+2\left( \frac{\dot{b}}{b}
\right)^{2}
\end{equation}
Now use (\ref{eq:4.16}), (\ref{eq:4.22}) and (\ref{eq:4.25}) which give $%
\frac{\dot{a}}{a}=-(tr K) - 2\frac{\dot{b}}{b}$, \mbox{$tr K
= -3H$}, and $\frac{\dot{b}}{b} = (\Sigma_{+}+1)H$ to express $\frac{\dot{a}%
}{a}$, $\frac{\dot{b}}{b}$ in (\ref{eq:4.31}) and obtain $\Omega = 1-
\Sigma_{+}^{2}$.

\item  ii) To obtain the first result in (\ref{eq:4.29}), use (\ref{eq:4.18}%
), (\ref{eq:4.19}), definition (\ref{eq:2.15}) of $T^{\alpha\beta}$, and (%
\ref{eq:4.5}) which gives $T_{22}=T_{33}$. Next, use
(\ref{eq:4.20}) to obtain \mbox{$0\leq R \leq$ 1} and
(\ref{eq:4.27}) to obtain $0\leq q \leq 2$ since $\Omega \geq 0 \,
\Rightarrow \, \Sigma_{+}^{2} \leq 1$.

\item  iii) Use expression (\ref{eq:2.3}) of $p^{0}$, expressions of $a^{2}$%
, $b^{2}$ in (\ref{eq:4.28}), the expressions of $P_{1}$, $P_{2}$ given by (%
\ref{eq:4.18}), (\ref{eq:4.19}) to obtain (4.30). This completes the proof of Lemma 4.1 $%
\blacksquare$
\end{itemize}

Now the system (\ref{eq:4.1})-(\ref{eq:4.2})-(\ref{eq:4.3}) writes, using
the notations (\ref{eq:4.17})-(\ref{eq:4.18})-(\ref{eq:4.19}):
\begin{equation}  \label{eq:4.32}
2\frac{\dot{a}\dot{b}}{ab} + \left( \frac{\dot{b}}{b} \right)^{2} = \rho
\end{equation}
\begin{equation}  \label{eq:4.33}
-\left[ 2\frac{\ddot{b}}{b} + \left( \frac{\dot{b}}{b} \right)^{2} \right] =
P_{1}
\end{equation}
\begin{equation}  \label{eq:4.34}
-\left[\frac{\ddot{a}}{a} + \frac{\ddot{b}}{b} + \frac{\dot{a}\dot{b}}{ab}
\right] = P_{2}
\end{equation}
It shows useful to explicit the second derivatives.
(\ref{eq:4.34}) gives, using (\ref {eq:4.33}) to express
$\frac{\ddot{b}}{b}$:
\begin{equation}  \label{eq:4.35}
\frac{\ddot{a}}{a}= -P_{2}+\frac{P_{1}}{2}+\frac{1}{2}\left( \frac{\dot{b}}{b%
} \right)^{2}-\frac{\dot{a}\dot{b}}{ab} = -P_{2}+\frac{P_{1}}{2}+\frac{1}{2}%
\left( \frac{\dot{b}}{b} \right)^{2}- \frac{2}{3}\frac{\dot{a}\dot{b}}{ab}-
\frac{1}{3}\frac{\dot{a}\dot{b}}{ab}
\end{equation}
and we write $\frac{\ddot{b}}{b}$ given by (4.33) on the form:
\begin{equation}  \label{eq:4.36}
\frac{\ddot{b}}{b}= -\frac{P_{1}}{2}- \frac{1}{2}\left( \frac{\dot{b}}{b}
\right)^{2} = -\frac{P_{1}}{2}- \frac{1}{3}\left( \frac{\dot{b}}{b}
\right)^{2}- \frac{1}{6}\left( \frac{\dot{b}}{b} \right)^{2}
\end{equation}
Now, if we use (\ref{eq:4.32}) to express the last terms in (\ref{eq:4.35})
and in (\ref{eq:4.36}), we obtain the Einstein evolution equations in the
form:
\begin{equation}  \label{eq:4.37}
\frac{\ddot{a}}{a}= \frac{2}{3}\left[ \left( \frac{\dot{b}}{b}
\right)^{2} - \frac{\dot{a}\dot{b}}{ab} \right]- \frac{\rho}{6} +
\frac{1}{2}(P_{1}-2P_{2})
\end{equation}
\begin{equation}  \label{eq:4.38}
\frac{\ddot{b}}{b}= \frac{1}{3}\left[ \frac{\dot{a}\dot{b}}{ab} -
\left( \frac{\dot{b}}{b} \right)^{2} \right]- \frac{\rho}{6}
-\frac{P_{1}}{2}
\end{equation}
We will also use the following relation, deduced from
(\ref{eq:4.16}), (\ref {eq:4.22}), (\ref{eq:4.25}):
\begin{equation}  \label{eq:4.39}
\frac{\dot{a}}{a} = (1-2\Sigma_{+})H
\end{equation}

\begin{remark}
\label{r:3.1} Assume $H>0$. Then: \newline
If $\dot{b}(t_{0}) = 0$ at a given time $t = t_{0}$, then the problem of the
global existence becomes trivial.
\end{remark}

The reason is the following: \newline (\ref{eq:4.36}) shows that
$\ddot{b} < 0$, then $\dot{b}$ is a decreasing function. By
(\ref{eq:4.28}) \newline \mbox{$(\Omega = 1-\Sigma_{+}^{2} \geq 0)
\, \Rightarrow \, (-1 \leq \Sigma_{+} \leq 1$)}; hence,
$\Sigma_{+} + 1 \geq 0$, and by (\ref{eq:4.25}), $\dot{b} \geq 0$.
But \mbox{($\dot{b} \geq 0, \dot{b}$ decreasing, $\dot{b}(t_{0})=0
)$ $\Rightarrow$ ( $\dot{b}(t)=0$ for $t \geq
t_{0}$ )}. Hence $b$ is constant and $b(t)=b(t_{0})$ for $t \geq t_{0}$.%
\newline
Then, by (\ref{eq:4.25}), $\Sigma_{+}(t) = -1$ for $t \geq t_{0}$, and by (%
\ref{eq:4.28}), $\Omega(t) =0$ for $t \geq t_{0}$; (\ref{eq:4.26}) then
implies $\rho(t) = 0$, for $t \geq t_{0}$, and since $0\leq P_{1}+2P_{2}
\leq \rho$, $P_{1}(t)=P_{2}(t)=0$ for $t \geq t_{0}$. The Einstein equations
(\ref{eq:4.32}), (\ref{eq:4.33}) are trivially satisfied for $t \geq t_{0}$
(since $\ddot{b}=0$), and by (\ref{eq:4.34}), $\ddot{a}(t)=0$ for $t \geq
t_{0}$. Then: $a(t)=C_{1}t + C_{2}$; $C_{1}, C_{2}$ constants.\newline
Notice that ($\rho(t)=0$ for $t \geq t_{0}$) $\Rightarrow$ ($f(t)=0$ for $t
\geq t_{0}$).\newline
\underline{Conclusion:} For $t \geq t_{0}$, the space-time becomes empty and
the problem of global existence becomes trivial.

In what follows, we will show that $H>0$ if $H(0)>0$, and we look for $b$
such that $\dot{b}>0$; consequently, by (\ref{eq:4.25}), $\Sigma_{+} + 1 >0$.

\begin{proposition}
\label{p:4.3} The Einstein evolution equations (\ref{eq:4.36})-(\ref{eq:4.37}%
) in $a$, $b$, are equivalent to the following non-linear first order system
in $(H, s, z, \Sigma_{+})$:

\[
\text{(IV .1)}\quad \left\{
\begin{array}{l}
\frac{dH}{dt}=-(1+q)H^{2}\qquad \qquad \qquad \qquad \qquad
\,\,\,\,\,\,\,\,\,\,\,\,\,\,\,\,\,\,\,\,\,\,\,\,\,\,\,\,\,\,\,\,\,\,\,\,\,\,%
\, \, \quad(4.40) \\
\frac{ds}{dt}=6s(1-s)\Sigma _{+}H\qquad \qquad \qquad \qquad \quad
\,\,\,\,\,\,\,\,\,\,\,\,\,\,\,\,\,\,\,\,\,\,\,\,\,\,\,\,\,\,\,\,\,\,\,\,\,\,%
\,\,\, \,\quad (4.41) \\
\frac{dz}{dt}=2z(1-z)(1+\Sigma _{+}-3s\Sigma _{+})H\qquad \quad
\,\,\,\,\,\,\,\,\,\,\,\,\,\,\,\,\,\,\,\,\,\,\,\,\,\,\,\,\,\,\,\,\,\,\,\,\,\,%
\,\, \, \quad(4.42) \\
\frac{d\Sigma _{+}}{dt}=-(2-q)\Sigma _{+}H+\Omega R_{+}H\qquad \quad \quad
\,\,\,\,\,\,\,\,\,\,\,\,\,\,\,\,\,\,\,\,\,\,\,\,\,\,\,\,\,\,\,\,\,\,\,\,\,\,%
\,\,\,\, \,\quad (4.43)
\end{array}
\right.
\]
\end{proposition}

\textbf{Proof:}

\begin{enumerate}
\item  \underline{Suppose we have (\ref{eq:4.37})-(\ref{eq:4.38})}.\newline

\begin{itemize}
\item  i) (\ref{eq:4.22}) and (\ref{eq:4.16}) give:
\[
\frac{dH}{dt}=\frac{1}{3} \left( \frac{\ddot{a}}{a}+2\frac{\ddot{b}}{b}%
-\left( \frac{\dot{a}}{a} \right)^{2}-2\left( \frac{\dot{b}}{b} \right)^{2}
\right)
\]
which gives, using (\ref{eq:4.37})-(\ref{eq:4.38}), (\ref{eq:4.31}) and (\ref
{eq:4.14}):
\[
\frac{dH}{dt}=\frac{1}{3}(\frac{\rho}{2}+\frac{P_{1}}{2}+P_{2}+\frac{1}{3}%
\left[(tr K)^{2 }-2\rho)\right].
\]
(4.40) then follows from (\ref{eq:4.20}),(\ref{eq:4.22}), (\ref{eq:4.26})
and (\ref{eq:4.27}).

\item  ii) Definition (4.24) of $s$ gives $1-s=\frac{2a^{2}}{b^{2}+2a^{2}}$
and we can write:
\[
\frac{ds}{dt}=2s(1-s)\frac{\dot{b}}{b}\left[1-\frac{\dot{a}}{a} \left( \frac{%
\dot{b}}{b} \right)^{-1} \right]
\]
we then deduce from (4.25) and (4.39) that:
\[
\frac{ds}{dt}=2s(1-s)(\Sigma_{+}+1)H\left( 1-\frac{(1-2\Sigma_{+})H}{%
(\Sigma_{+}+1)H} \right)=6s(1-s)\Sigma_{+}H
\]

\item  iii) Definition (4.23) of $z$ gives:
\[
\frac{dz}{dt}=2z^{2} \left( \frac{\dot{a}}{a}\frac{1}{a^{2}}+2 \frac{\dot{b}%
}{b}\frac{1}{b^{2}} \right).
\]
(4.42) then follows from (4.39), (4.25) and (4.28).

\item  iv) We can write, using (4.25) and (4.40):
\begin{equation}
\frac{d\Sigma_{+}}{dt}=\frac{d}{dt}\left( \frac{1}{H}\frac{\dot{b}}{b}
\right)=\frac{\dot{b}}{b}(1+q)+\frac{1}{H}\left[ \frac{\ddot{b}}{b}- \left(
\frac{\dot{b}}b{} \right)^{2} \right]  \tag{a}
\end{equation}
Now the evolution equation (4.38) gives, using (4.25) and (4.39)
\begin{align*}
\frac{\ddot{b}}{b}-\left( \frac{\dot{b}}b{} \right)^{2}&=\frac{\dot{b}}{b}
\left( \frac{1}{3}\frac{\dot{a}}{a}-\frac{4}{3}\frac{\dot{b}}{b} \right) -%
\frac{\rho}{6}-\frac{P_{1}}{2}&=\frac{\dot{b}}{b}(-1-2\Sigma_{+})H -\frac{%
\rho}{6}-\frac{P_{1}}{2}
\end{align*}
so that, (a) gives, using once more (4.25):
\begin{align*}
\frac{d\Sigma_{+}}{dt}&=\frac{\dot{b}}{b}(q-2\Sigma_{+})-\frac{\rho}{6H}-%
\frac{P_{1}}{2H} =(\Sigma_{+}+1)(q-2\Sigma_{+})H-\frac{\rho}{6H}-\frac{P_{1}%
}{2H} \\
&=(q-2)\Sigma_{+}H-2\Sigma_{+}^{2}H+qH-\frac{\rho}{6H}-\frac{P_{1}}{2H}
\end{align*}
Now express $q$ in the third term by (4.27), use (4.26) that gives $\frac{1}{%
H}=\frac{3\Omega H}{\rho}$, and definitions (4.20) and (4.21) of $R$ and $%
R_{+}$ to obtain (4.43).
\end{itemize}

\item  \underline{Suppose we have (4.40)-(4.41)-(4.42)-(4.43)}.

\begin{itemize}
\item  i) Define the quantities $a^{2}$, $b^{2}$ as in (4.28) and the other
quantities the same way as we did. Differentiating both sides of the
corresponding formula (4.39) yields:
\[
\frac{\ddot{a}}{a}-\left( \frac{\dot{a}}{a} \right)^{2}=\dot{H}%
(1-2\Sigma_{+})-2H\dot{\Sigma}_{+}
\]
which gives, using (4.40) and (4.41):
\begin{align*}
\frac{\ddot{a}}{a}&=-(1+q)(1-2\Sigma_{+})H^{2}-2H^{2}[(q-2)\Sigma_{+}+\Omega
R_{+}]+\left( \frac{\dot{a}}{a} \right)^{2} \\
&= -H^{2}+6\Sigma_{+}H^{2}+\left( \frac{\dot{a}}{a} \right)^{2}-qH^{2}-2%
\Omega R_{+}H^{2}
\end{align*}
Now, use expression (4.27) of $q$ to obtain
\begin{equation}
\frac{\ddot{a}}{a}=\left(-H^{2}+6\Sigma_{+}H^{2}-2\Sigma_{+}^{2}H^{2}+
\left( \frac{\dot{a}}{a} \right)^{2}\right)- \left[ \frac{\Omega H^{2}}{2}
(1+R)+2\Omega H^{2}R_{+} \right]  \tag{b}
\end{equation}
For the second term of (b), use (4.26) which gives $\Omega H^{2}=\frac{\rho}{%
3}$ and the definitions (4.20) and (4.21) of $R$ and $R_{+}$ to obtain:
\[
-\frac{\Omega H^{2}}{2}(1+R)-2\Omega H^{2}R_{+}=-\frac{\rho}{6}+\frac{1}{2}%
(P_{1}-2P_{2})
\]
Next, for the first term of (b), use (4.25) to express $\Sigma_{+}$, (4.16)
and (4.22) that give $H=\frac{1}{3}(\frac{\dot{a}}{a}+2\frac{\dot{b}}{b})$
to obtain:
\[
-H^{2}+6\Sigma_{+}H^{2}-2\Sigma_{+}^{2}H^{2}+ \left( \frac{\dot{a}}{a}
\right)^{2}=\frac{2}{3}\left[ \left( \frac{\dot{b}}{b} \right)^{2}-\frac{%
\dot{a}\dot{b}}{ab} \right]
\]
and (4.37) follows.

\item  ii) Differentiating both sides of $\frac{\dot{b}}{b}=(\Sigma_{+}+1)H$
yields:
\[
\frac{\ddot{b}}{b} - \left( \frac{\dot{b}}{b} \right)^{2}=\dot{\Sigma}%
_{+}H+(\Sigma_{+}+1)\dot{H}
\]
which gives, using (4.40) and (4.43) and simplifying:
\[
\frac{\ddot{b}}{b} = (-3\Sigma_{+}-1)H^{2}-qH^{2}+\Omega R_{+}H^{2}+ \left(
\frac{\dot{b}}{b} \right)^{2}
\]
Now this writes using definition (4.27) of $q$:
\[
\frac{\ddot{b}}{b}=\left[ (-2\Sigma_{+}^{2}-3\Sigma_{+}-1)H^{2}+ \left(
\frac{\dot{b}}{b} \right)^{2} \right]-\frac{\Omega H^{2}}{2}(1+R)+\Omega
H^{2}R_{+} \tag{c}
\]
For the second term of (c), use (4.26) which gives $\Omega H^{2}=\frac{\rho}{%
3}$ and definitions (4.20) and (4.21) of $R$ and $R_{+}$ to obtain:
\[
-\frac{\Omega H^{2}}{2}(1+R)+\Omega H^{2}R_{+}=-\frac{\rho}{6}-\frac{P_{1}}{2%
}
\]
Next, for the first term of (c), use (4.25) to express $\Sigma_{+}$ and \mbox{$H=%
\frac{1}{3}\left(\frac{\dot{a}}{a}+2\frac{\dot{b}}{b} \right)$} to
obtain:
\[
(-2\Sigma_{+}^{2}-3\Sigma_{+}-1)H^{2}+ \left( \frac{\dot{b}}{b} \right)^{2}=%
\frac{1}{3}\left[ \frac{\dot{a}\dot{b}}{ab}-\left( \frac{\dot{b}}{b}
\right)^{2} \right]
\]
and (4.38) follows. This completes the proof of Proposition 4.2 $%
\blacksquare $
\end{itemize}
\end{enumerate}

Now observe, using expressions (4.20), (4.21) and (4.27) of $R$, $R_{+}$ and
$q$, that these quantities express in terms of $\frac{1}{\rho}$ and we have
nothing to bound such quantities. This is why, we give a new formulation of
equations (4.40) and (4.43), that raise this problem, as follows:

\begin{proposition}\label{p:4.4}
 Equations (4.40) and (4.43) can write respectively:
\begin{equation}
\frac{dH}{dt}=-\frac{3}{2}(1+\Sigma_{+}^{2})H^{2}-\frac{P_{1}+2P_{2}}{6}
\tag{4.44}
\end{equation}
\begin{equation}
\frac{d\Sigma_{+}}{dt}=-\frac{3}{2}(1-\Sigma_{+}^{2})H\Sigma_{+}+\frac{P_{1}%
}{6H} (\Sigma_{+}-2)+\frac{P_{2}}{3H} (\Sigma_{+}+1)  \tag{4.45}
\end{equation}
\end{proposition}

\textbf{Proof:}

\begin{itemize}
\item  i) Write (4.40) using definition (4.27) of $q$ in which $R$ is given
by (4.20); next, use (4.28) and (4.26) which give $\Omega =1-\Sigma_{+}^{2}$
and $\frac{\Omega H^{2}}{\rho}=\frac{1}{3}$ to obtain (4.44).

\item  ii) Write (4.43) using definition (4.27) of $q$ and apply definition
(4.20) and (4.21) of $R$ and $R_{+}$; next use once more (4.28) and (4.26)
which give $\Omega =1-\Sigma_{+}^{2}$ and $\frac{\Omega H}{\rho}=\frac{1}{3H}
$ to obtain (4.45) $\blacksquare$
\end{itemize}

By prop.\ref{p:4.4} system (IV.1) is then equivalent to:
\[
\,\,\,\,\text{(IV.2)}\,\,\,\,\,\,\,\,\,\,\,\,\,\,\,\,\,\,\left\{
\begin{array}{l}
\frac{dH}{dt}=-\frac{3}{2}(1+\Sigma _{+}^{2})H^{2}-\frac{P_{1}+2P_{2}}{6}%
\qquad \qquad \qquad \qquad \qquad \,\,\,\,\,\,\,\text{(4.44)} \\
\frac{ds}{dt}=6s(1-s)\Sigma _{+}H\qquad
\,\,\,\,\,\,\,\,\,\,\,\,\,\,\,\,\,\,\,\,\,\,\,\,\,\,\,\,\,\,\,\,\,\,\,\,\,\,%
\,\,\,\,\,\,\,\,\,\,\,\,\,\,\,\,\,\,\,\,\,\,\,\,\,\,\,\,\,\,\,\,\,\,\,\,\,\,%
\,\,\,\,\,\,\,\,\,\text{(4.41)} \\
\frac{dz}{dt}=2z(1-z)(1+\Sigma _{+}-3s\Sigma _{+})H\qquad \quad \qquad
\qquad \qquad \,\,\,\,\,\,\text{(4.42)} \\
\frac{d\Sigma _{+}}{dt}=-\frac{3}{2}(1-\Sigma _{+}^{2})H\Sigma _{+}+\frac{%
P_{1}}{6H}(\Sigma _{+}-2)+\frac{P_{2}}{3H}(\Sigma _{+}+1)\,.\,\,\,\,\,\text{%
(4.45)}
\end{array}
\right.
\]

It shows useful in what follows to introduce the following notation.

\begin{definition}
\label{d:4.5} Let $0<s<1$. Set:
\begin{equation}
\alpha (s)=Inf(s, 1-s)  \tag{4.46}
\end{equation}
\end{definition}

For the global existence theorem we will need the following a priori
estimations.

\begin{proposition}
\label{p:4.6} Let $\delta >0$ be given and let $t_{0} \in \mathbb{R}_{+}$.
Then any solution $(H, s, z, \Sigma_{+})$ of the system (IV.2) on $[t_{0},
t_{0}+\delta]$, with $H(t_{0}) >0$, satisfy the following a priori
estimations, $\forall t \in [0, \delta]$:
\begin{equation}
H(t_{0}+t)=\frac{H(t_{0})}{1+H(t_{0}) \int_{t_{0}}^{t_{0}+t}(1+q)d\tau}
\tag{4.47}
\end{equation}
\begin{equation}
\frac{1}{H(t_{0}+t)} \leq \frac{1}{H(t_{0})}e^{3H(t_{0})t}  \tag{4.48}
\end{equation}
\begin{equation}
\frac{1}{\alpha(s(t_{0}+t))} \leq \frac{1}{\alpha(s(t_{0}))}e^{6H(t_{0})t}
\tag{4.49}
\end{equation}
\begin{equation}
\frac{1}{\alpha(z(t_{0}+t))} \leq \frac{1}{\alpha(z(t_{0}))}e^{10H(t_{0})t}
\tag{4.50}
\end{equation}
\begin{equation}
-1 < \Sigma_{+}(t_{0}+t) \leq \frac{1}{2}  \tag{4.51}
\end{equation}
\end{proposition}

\textbf{Proof:} Let $t \in [0, \delta]$. By prop.\ref{p:4.4}, for $H$ we
consider (4.40) or (4.44), and for $\Sigma_{+}$ we consider (4.43) or (4.45).

\begin{enumerate}
\item  (4.40) writes: $-\frac{dH}{H^{2}}=(1+q)dt$; integrating
both sides on $[t_{0}, t_{0}+t]$ yields (4.47). Notice that, using
(4.40) or (4.44), $H$ is a decreasing function.

\item  By (4.29) we have $0 \leq q \leq 2$; so, by (4.47), $(H(t_{0}) >0)$ $%
\Rightarrow$ $(H(t_{0}+t) >0)$ and $0<H(t_{0}+t) \leq H(t_{0})$. Now (4.40)
also writes: $-\frac{dH}{H}=(1+q)Hdt$; integrating both sides on $[t_{0},
t_{0}+t]$ and using $0 \leq q \leq 2$, $H(t_{0}+t) \leq H(t_{0})$ yields
(4.48).

\item  (4.41) gives:
\[
\left| \frac{ds}{s} \right|=|6(1-s)\Sigma_{+}H|dt; \quad \left| \frac{d(1-s)%
}{1-s} \right|= |6s\Sigma_{+}H|dt
\]
Now integrate on $[t_{0}, t_{0}+t]$, using $|\Sigma_{+}| \leq 1$, $0<s<1$, $%
0< H \leq H(t_{0})$ to obtain:
\[
\left\{
\begin{array}{ll}
e^{-6H(t_{0})t} \leq \frac{s(t_{0}+t)}{s(t_{0})} \leq e^{6H(t_{0})t} &  \\
e^{-6H(t_{0})t} \leq \frac{1-s(t_{0}+t)}{1-s(t_{0})} \leq e^{6H(t_{0})t} &
\end{array}
\right.
\]
We then deduce, using definition (4.46) of $\alpha(s)$ that:
\[
\alpha(s(t_{0}))e^{-6H(t_{0})t} \leq \alpha(s(t_{0}+t))
\]
and (4.49) follows.

\item  To obtain (4.50), proceed as above, using this time (4.42) which
gives:
\[
\left| \frac{dz}{z} \right|=|2(1-z)(1+\Sigma_{+}-3s\Sigma_{+})H|dt; \quad
\left| \frac{d(1-z)}{1-z} \right|=|2z(1+\Sigma_{+}-3s\Sigma_{+})H|dt
\]
Notice that the inequalities (4.48), (4.49) and (4.50) don't involve $f$,
because we used (4.40) in which $0 \leq q \leq 2$ and (4.41), (4.42), which
contain no source terms of the Einstein equations.

\item  To prove (4.51), we show that, for every point $t_{1} \in [t_{0},
t_{0}+\delta]$,
\begin{equation}
(\Sigma_{+}(t_{1})=\frac{1}{2}) \quad \Rightarrow \quad (\dot{\Sigma}%
_{+}(t_{1}) \leq 0)  \tag{a}
\end{equation}
Since (a) will shows that $\frac{1}{2}$ is an upper bound for all the values
$\Sigma_{+}(t)$, $t \in [t_{0}, t_{0}+\delta]$. We use equation (4.45) in $%
\Sigma_{+}$. Evaluating both sides of (4.45) at $t=t_{1}$ gives, using $%
\Sigma_{+}(t_{1})=\frac{1}{2}$:
\begin{equation}
\dot{\Sigma}_{+}(t_{1})=\frac{1}{H}(-\frac{9}{16}H^{2}-\frac{P_{1}}{4}+\frac{%
P_{2}}{2})  \tag{b}
\end{equation}
Now, the relation $\Omega =1-\Sigma_{+}^{2}$ gives $\Omega(t_{1})=1-\frac{1}{%
4}=\frac{3}{4}$. We then deduce from (4.26) that, for $t=t_{1}$, $H^{2}=%
\frac{4\rho}{9}$, and (b) gives:
\begin{equation}
\dot{\Sigma}_{+}(t_{1})=\frac{1}{H}\left(\frac{-\rho -P_{1}
+2P_{2}}{4}\right)  \tag{c}
\end{equation}
Now, the inequalities $0 \leq P_{1}+2P_{2} \leq \rho$ in (4.29) give : $%
2P_{2} \leq \rho -P_{1}$ so that (c) implies:
\begin{equation}
\dot{\Sigma}_{+}(t_{1})=\frac{1}{H}(-\frac{P_{1}}{2}) \leq 0  \tag{c}
\end{equation}
since $P_{1} \geq 0$; and (a) follows. This completes the proof of
proposition \ref{p:4.6} $\blacksquare$\newline
\end{enumerate}

We deduce:

\begin{proposition}
\label{p:4.7} Let $f \in C(0, T; X_{r})$, $T>0$, be given. Suppose
the initial value problem for the system (IV.2), with initial data
$(H_{0},
s_{0}, z_{0}, \Sigma_{+0})$, with $H_{0}>0$, $0<s_{0}<1$, $0<z_{0}<1$, $%
-1<\Sigma_{+0} \leq \frac{1}{2}$, at $t=0$, has a solution $(\breve{H},%
\breve{s},\breve{z},\breve{\Sigma}_{+})$ on $[0, t_{0}]$ with $0
\leq t_{0} <T$. Then, any solution $(H, s, z, \Sigma_{+})$ of the
initial value problem (IV.2) on $[t_{0}, t_{0}+\delta]$, $\delta >0$, with initial data $%
(H, s, z, \Sigma_{+})(t_{0}) = (\breve{H},\breve{s},\breve{z},\breve{\Sigma}%
_{+})(t_{0})$, at $t=t_{0}$, satisfy, for $t \in [0, \delta]$, the
inequalities:
\begin{equation}
\frac{1}{H(t_{0}+t)} \leq \gamma_{0}e^{3H_{0}(T+\delta)}  \tag{4.52}
\end{equation}
\begin{equation}
\frac{1}{\alpha(s(t_{0}+t)(} \leq \gamma_{0}e^{6H_{0}(T+\delta)}  \tag{4.53}
\end{equation}
\begin{equation}
\frac{1}{\alpha(z(t_{0}+t)(} \leq \gamma_{0}e^{10H_{0}(T+\delta)}  \tag{4.54}
\end{equation}
where
\begin{equation}
\gamma_{0} =\left( \frac{1}{H_{0}} + \frac{1}{s_{0}} + \frac{1}{z_{0}}
\right)  \tag{4.55}
\end{equation}
\end{proposition}

\textbf{Proof:} The hypothesis on $f$ implies that system (IV.2)
is defined on $[0,T]$. Apply prop.\ref{p:4.6}. Notice that by
(4.40), $\breve{H}$ is a decreasing function; so (4.48), (4.49),
(4.50) give in the present case, since $\breve{H}(t_{0}) \leq
H_{0}$:
\begin{align*}
\begin{cases} \frac{1}{H(t_{0}+t)} \leq
\frac{1}{\breve{H}(t_{0})}e^{3H_{0}t}; \quad \frac{1}{\alpha(s(t_{0}+t))}
\leq \frac{1}{\alpha(\breve{s}(t_{0}))}e^{6H_{0}t}\\
\,\frac{1}{\alpha(z(t_{0}+t))} \leq
\frac{1}{\alpha(\breve{z}(t_{0}))}e^{10H_{0}t} \tag{a} \end{cases}\, t\in
[0, \delta]
\end{align*}
Now apply prop.\ref{p:4.6} to the solution $(\breve{H},\breve{s},\breve{z},%
\breve{\Sigma}_{+})$ of (IV.2) on $[0, t_{0}]$, with \newline
\mbox{$(\breve{H},\breve{s},\breve{z},\breve{\Sigma}_{+})(0)=(H_{0},
s_{0}, z_{0}, \Sigma_{+0})$}; then (4.48), (4.49), (4.50) in which we take $%
t_{0}=0$ and $t=t_{0}$ give:
\begin{equation}
\frac{1}{\breve{H}(t_{0})} \leq \frac{1}{H_{0}}e^{3H_{0}t_{0}}; \,\frac{1}{%
\alpha(\breve{s}(t_{0}))} \leq \frac{1}{\alpha(s_{0})}e^{6H_{0}t_{0}}; \,%
\frac{1}{\alpha(\breve{z}(t_{0}))} \leq \frac{1}{\alpha(z_{0})}%
e^{10H_{0}t_{0}}  \tag{b}
\end{equation}
(4.52), (4.53), (4.54) then follow from (a), (b), $t_{0}<T$, $t \leq \delta$
and the definition (4.55) of $\gamma_{0}$. This completes the proof of prop.%
\ref{p:4.7} $\blacksquare$

\begin{corollary}\label{c:4.8} Any solutions $a$, $b$ of the Einstein equations
(4.1)-(4.2)-(4.3) on $[t_{0}, t_{0}+ \delta]$, $t_{0} \in \mathbb{R_{+}}$, $%
\delta \geq 0$, are increasing functions.
\end{corollary}

\textbf{Proof:} (4.51) shows that $1-2\Sigma_{+} \geq 0$ on $[t_{0}, t_{0}+
\delta]$; then (4.39) implies $\dot{a}>0$, since $a>0$ and $H>0$; and $a$ is
an increasing function. Next, (4.25) gives: $\frac{\dot{b}}{b}%
=(\Sigma_{+}+1)H$ and since $\Sigma_{+}+1 \geq 0$, $H \geq 0$, $b>0$ this
implies $\dot{b}>0$ and $b$ is an increasing function $\blacksquare$

\begin{remark}
\label{r:4.9} By corollary \ref{c:4.8} if we take in (4.4) the initial data $%
a_{0}$, $b_{0}$, such that
\[
a_{0} \geq \frac{3}{2}; \quad b_{0} \geq \frac{3}{2}
\]
then $a$ and $b$ satisfy the hypothesis of the existence theorem \ref{t:3.2}
of solutions for the Boltzmann equation.
\end{remark}

We now prove the existence theorem for the Einstein equations, which are
equivalent, using prop.\ref{p:4.4}, to the system (IV.2). Following Prop.\ref
{p:4.6}, we will take the initial datum $H_{0}$ for $H$ such that $H_{0}>0$,
and denote the initial data for the system (IV.2) at $t=0$, $(H_{0}, s_{0},
z_{0}, \Sigma_{+0})$, where:
\begin{equation}
H_{0}>0;\quad 0<s_{0}<1; \quad 0<z_{0}<1; \quad -1<\Sigma_{+0}\leq \frac{1}{2%
}  \tag{4.56}
\end{equation}
We will apply the standard theory on the first order differential system.
For this purpose, we will have to study the function defined by the r.h.s of
the system (IV.2), i.e, the function $F$ defined by:
\begin{equation}
F(t,\, H,\,s,\,z,\,\Sigma _{+})=\left\{
\begin{array}{l}
-\frac{3}{2}(1+\Sigma _{+}^{2})H^{2}-\frac{P_{1}+2P_{2}}{6} \\
6s(1-s)\Sigma _{+}H\qquad \qquad \\
2z(1-z)(1+\Sigma _{+}-3s\Sigma _{+})H \\
-\frac{3}{2}(1-\Sigma _{+}^{2})H\Sigma _{+}+\frac{P_{1}}{6H}(\Sigma _{+}-2)+%
\frac{P_{2}}{3H}(\Sigma _{+}+1)
\end{array}
\right.  \tag*{(4.57)}
\end{equation}
(4.40) shows that $H$ is a decreasing function, so that $H(t) \leq H_{0}$, $%
t \geq 0$, whenever $H$ exists. We are then funded to suppose, using (4.28),
(4.51), that $F$ is defined for:
\begin{equation}
(H, s, z, \Sigma_{+}) \in B=]0, H_{0}]\times ]0, 1[ \times ]0,1[ \times ]-1,
\frac{1}{2}]  \tag{4.58}
\end{equation}
We will have to prove that $F$ is continuous in $t$ and locally
Lipschitzian in $X=(H, s, z, \Sigma_{+})$, with respect to the
norm of $\mathbb{R}^{4}$ we take to be\\ \mbox{$\| X \|=
|H|+|s|+|z|+|\Sigma_{+}|$}. $F$ depends on $t$ through $f$ in
$P_{1}$ and $P_{2}$ (see their expressions in (4.30)); but, by
hypothesis, $f$ is continuous in $t$ and so is $F$.

A glance to (4.57) shows that the real problem will be to prove that $P_{1}$
and $P_{2}$ defined in (4.30) are locally Lipschitzian .\newline
We then begin by proving:

\begin{lemma}
\label{l:4.10} Suppose $f \in C\left([t_{0}, t_{0}+\delta]; L^{1}_{2}(%
\mathbb{R}^{3})\right)$, $t_{0} \geq 0$, $\delta >0$.\newline
Let $s, s_{1}, s_{2}, z, z_{1}, z_{2} \in ]0, 1[$, $H_{1}, H_{2} \in ]0,
H_{0}]$, $t \in [t_{0}, t_{0}+ \delta]$. then:
\begin{equation}
\left| P_{i}(s_{1}, z_{1}, f)- P_{i}(s_{2}, z_{2}, f) \right|\leq \frac{C \|
f(t) \|(|s_{1} -s_{2}| +|z_{1}-z_{2}|)}{\alpha^{6}(s_{1})\alpha^{5}(s_{2})%
\alpha^{8}(z_{1}) \alpha^{5}(z_{2})} ; \, i=1, 2  \tag{4.59}
\end{equation}
\begin{equation}
|P_{i}(s, z, f)| \leq \frac{C \| f(t) \|}{\alpha^{3}(s)\alpha^{3}(z)}; \quad
i=1, 2  \tag{4.60}
\end{equation}
\begin{equation}
\left|\frac{ P_{i}(s_{1}, z_{1}, f)}{H_{1}}- \frac{P_{i}(s_{2}, z_{2},f)}{%
H_{2}}\right|\leq\frac{C\|f(t)\|(|s_{1}-s_{2}|+|z_{1}-z_{2}|+|H_{1}-H_{2}|)%
} {H_{2}(1+H_{1})\alpha^{6}(s_{1})\alpha^{5}(s_{2})\alpha^{8}(z_{1})
\alpha^{5}(z_{2})}; \, i=1, 2  \tag{4.61}
\end{equation}
where $C>0$ is a constant.
\end{lemma}

\textbf{Proof:} Consider the expressions of $p^{0}, P_{1}, P_{2}$ in (4.30).
Let us begin by bounding the differences involved in $P_{i}(s_{1}, z_{1},
f)- P_{i}(s_{2}, z_{2}, f), i=1, 2$. We have:
\begin{align*}
\frac{1}{p^{0}(s_{1}, z_{1})}-\frac{1}{p^{0}(s_{2}, z_{2})}=\frac{\left(
\frac{z_{2}} {s_{2}(1-z_{2})}-\frac{z_{1}} {s_{1}(1-z_{1})}
\right)(p^{1})^{2}}{p^{0}(s_{1},z_{1})p^{0}(s_{2},z_{2})\left( p^{0}(s_{1},
z_{1})+ p^{0}(s_{2}, z_{2})\right)} + \\
\frac{\left(\frac{2z_{2}} {(1-s_{2})(1-z_{2})}-\frac{2z_{1}}{%
(1-s_{1})(1-z_{1})} \right)\left((p^{2})^{2}+ (p^{3})^{2}\right)}{%
p^{0}(s_{1},z_{1})p^{0}(s_{2},z_{2})\left( p^{0}(s_{1}, z_{1})+ p^{0}(s_{2},
z_{2})\right)}  \tag{a}
\end{align*}
But $p^{0}(s_{1},z_{1})p^{0}(s_{2},z_{2})\left( p^{0}(s_{1},z_{1})+
p^{0}(s_{2}, z_{2})\right) \geq (p^{0}(s_{1},z_{1}))^{2}p^{0}(s_{2},z_{2})$
and (4.30) shows that $(p^{0}(s_{1},z_{1}))^{2}p^{0}(s_{2},z_{2})$ is
bounded from below by each one of the following four quantities, that will
serve to bound $P_{1}$ and $P_{2}$ respectively:
\[
\begin{cases}
\frac{z_{1}}{s_{1}(1-z_{1})}\left(\frac{z_{2}}{s_{2}(1-z_{2})}\right)^{%
\frac{1}{2}} |p^{1}|^{3} \\
\frac{2z_{1}}{(1-s_{1})(1-z_{1})}\left(\frac{z_{2}} {s_{2}(1-z_{2})}
\right)^{\frac{1}{2}}|p^{1}|((p^{2})^{2}+(p^{3})^{2}) \end{cases}  \tag{b}
\]
\[
\begin{cases}
\frac{z_{1}}{s_{1}(1-z_{1})}\left(\frac{2z_{2}}{(1-s_{2})(1-z_{2})}\right)^{%
\frac{1}{2}} (p^{1})^{2}|p^{2}| \\
\frac{2z_{1}}{(1-s_{1})(1-z_{1})}\left(\frac{2z_{2}} {(1-s_{2})(1-z_{2})}
\right)^{\frac{1}{2}}|p^{2}|((p^{2})^{2}+(p^{3})^{2}) \end{cases}  \tag{c}
\]
quantities (b) will serve for $P_{1}$ and (c) for $P_{2}$. Next, we have:
\begin{equation}
\begin{cases} \,\left| \frac{z_{2}} {s_{2}(1-z_{2})}-\frac{z_{1}}
{s_{1}(1-z_{1})}\right| \leq
\frac{C(|s_{1}-s_{2}|+|z_{1}-z_{2}|)}{s_{1}s_{2}(1-z_{1})(1-z_{2})}\\
\left|\frac{2z_{2}}{(1-s_{2})(1-z_{2})}-\frac{2z_{1}}{(1-s_{1})(1-z_{1})}%
\right| \leq
\frac{C(|s_{1}-s_{2}|+|z_{1}-z_{2}|)}{(1-s_{1})(1-s_{2})(1-z_{1})(1-z_{2})}
\end{cases}  \tag{d}
\end{equation}
(a) then gives, using (b)-(d) and (c)-(d) respectively:
\[
\left| \frac{1}{p^{0}(s_{1}, z_{1})}-\frac{1}{p^{0}(s_{2}, z_{2})}\right|
\leq \frac{C(|s_{1}-s_{2}|+|z_{1}-z_{2}|)}{%
s_{1}s_{2}(1-s_{1})(1-s_{2})z_{1}z_{2}^{\frac{1}{2}}}.\frac{1}{|p^{1}|} %
\tag{e}
\]
\[
\left| \frac{1}{p^{0}(s_{1}, z_{1})}-\frac{1}{p^{0}(s_{2}, z_{2})}\right|
\leq \frac{C(|s_{1}-s_{2}|+|z_{1}-z_{2}|)}{%
s_{1}s_{2}(1-s_{1})(1-s_{2})z_{1}z_{2}^{\frac{1}{2}}}.\frac{1}{|p^{2}|} %
\tag{f}
\]
We now bound the differences in the coefficients of the integrals in $P_{1}$%
, $P_{2}$ which are given by (4.30). Concerning $P_{1}$, we have:
\begin{align*}
&\frac{1}{s_{1}^{\frac{3}{2}}(1-s_{1})}\left(\frac{z_{1}}{1-z_{1}}\right)^{%
\frac{5}{2}}- \frac{1}{s_{2}^{\frac{3}{2}}(1-s_{2})}\left(\frac{z_{2}}{%
1-z_{2}}\right)^{\frac{5}{2}}= \\
&\left( \frac{1}{s_{1}^{\frac{3}{2}}(1-s_{1})}-\frac{1}{s_{2}^{\frac{3}{2}%
}(1-s_{2})} \right)\left(\frac{z_{1}}{1-z_{1}}\right)^{\frac{5}{2}} +\frac{1%
}{s_{2}^{\frac{3}{2}}(1-s_{2})}\left[ \left(\frac{z_{1}}{1-z_{1}}\right)^{%
\frac{5}{2}}- \left(\frac{z_{2}}{1-z_{2}}\right)^{\frac{5}{2}}\right]
\tag{g}
\end{align*}
in which:
\begin{align*}
& \left| \frac{1}{s_{1}^{\frac{3}{2}}(1-s_{1})}-\frac{1}{s_{2}^{\frac{3}{2}%
}(1-s_{2})} \right|=\left| \frac{s_{2}^{\frac{3}{2}}(1-s_{2})-s_{1}^{\frac{3%
}{2}}(1-s_{1})} {(s_{2}s_{1})^{\frac{3}{2}}(1-s_{1})(1-s_{2})} \right|= \\
&\left| \frac{(\sqrt{s_{2}^{3}}-\sqrt{s_{1}^{3}})(1-s_{2})+s_{1}^{\frac{3}{2}%
}(s_{1}- s_{2})} {(s_{2}s_{1})^{\frac{3}{2}}(1-s_{1})(1-s_{2})} \right| \leq
\frac{C|s_{1}-s_{2}|}{s_{1}^{3}s_{2}^{\frac{3}{2}}(1-s_{1})(1-s_{2})}
\end{align*}
and:
\begin{align*}
&\left| \left(\frac{z_{1}}{1-z_{1}}\right)^{\frac{5}{2}}- \left(\frac{z_{2}}{%
1-z_{2}}\right)^{\frac{5}{2}} \right|= \left| \sqrt{\left(\frac{z_{1}}{%
1-z_{1}}\right)^{5}}-\sqrt{\left(\frac{z_{2}}{1-z_{2}}\right)^{5}} \right|
\leq \\
& \frac{C\left| \left(\frac{z_{1}}{1-z_{1}}\right)^{5}-\left(\frac{z_{2}}{%
1-z_{2}}\right)^{5} \right|}{z_{1}^{\frac{5}{2}}} \leq \frac{C|z_{1}-z_{2}|}{%
z_{1}^{\frac{5}{2}}(1-z_{1})^{5} (1-z_{2})^{5}}
\end{align*}
so that (g) gives:
\begin{align*}
\left| \frac{1}{s_{1}^{\frac{3}{2}}(1-s_{1})}\left(\frac{z_{1}}{1-z_{1}}%
\right)^{\frac{5}{2}}- \frac{1}{s_{2}^{\frac{3}{2}}(1-s_{2})}\left(\frac{%
z_{2}}{1-z_{2}}\right)^{\frac{5}{2}} \right| \leq \\
\frac{C(|s_{1}-s_{2}|+|z_{1}-z_{2}|)}{s_{1}^{3}(1-s_{1})s_{2}^{\frac{3}{2}%
}(1-s_{2})z_{1}^{\frac{5}{2}} (1-z_{1})^{5}(1-z_{2})^{5}}  \tag{h}
\end{align*}
An analogous calculation gives, for the coefficient of the integral in $%
P_{2} $
\begin{align*}
\left| \frac{1}{s_{1}^{\frac{1}{2}}(1-s_{1})^{2}}\left(\frac{z_{1}}{1-z_{1}}%
\right)^{\frac{5}{2}}- \frac{1}{s_{2}^{\frac{1}{2}}(1-s_{2})^{2}}\left(\frac{%
z_{2}}{1-z_{2}}\right)^{\frac{5}{2}} \right| \leq \\
\frac{C(|s_{1}-s_{2}|+|z_{1}-z_{2}|)}{s_{1}(1-s_{1})^{2}s_{2}^{\frac{1}{2}%
}(1-s_{2})^{2} z_{1}^{\frac{5}{2}}(1-z_{1})^{5}(1-z_{2})^{5}}  \tag{i}
\end{align*}
We can now establish (4.59). We have, using the expression of $P_{1}$ in
(4.30)
\begin{align*}
&P_{1}(s_{1}, z_{1}, f)-P_{1}(s_{2}, z_{2}, f)=16\pi\left[\frac{1}{s_{1}^{%
\frac{3}{2}}(1-s_{1})}\left(\frac{z_{1}}{1-z_{1}}\right)^ {\frac{5}{2}}-
\frac{1}{s_{2}^{\frac{3}{2}}(1-s_{2})}\left(\frac{z_{2}}{1-z_{2}}\right)^{%
\frac{5}{2}}\right] \times \\
&\int_{\mathbb{R}^{3}}\frac{f(p^{1})^{2}d\bar{p}}{p^{0}(s_{1}, z_{1})}+
\frac{16\pi}{s_{2}^{\frac{3}{2}}(1-s_{2})}\left(\frac{z_{2}}{1-z_{2}}%
\right)^{\frac{5}{2}} \int_{\mathbb{R}^{3}}f(p^{1})^{2}\left(\frac{1}{%
p^{0}(s_{1}, z_{1})}-\frac{1}{p^{0}(s_{2}, z_{2})}\right)d\bar{p}
\end{align*}
Notice that:
\[
p^{0}(s_{1}, z_{1})>\left(\frac{z_{1}}{s_{1}(1-z_{1})}\right)^{\frac{1}{2}%
}|p^{1}| \tag{j}
\]
(4.59) for $P_{1}$ then follows from (h), (e), (j), $|p^{1}|<\sqrt{1+|\bar{p}%
|^{2}}$, the definition (4.46) of $\alpha(s)$, $0<s<1$, and $0<\alpha(s)<1.$

Now, concerning $P_{2}$, we have, using its expression in (4.30):
\begin{align*}
&P_{2}(s_{1}, z_{1}, f)-P_{2}(s_{2}, z_{2}, f)=32\pi\left[\frac{1}{s_{1}^{%
\frac{1}{2}}(1-s_{1})^{2}}\left(\frac{z_{1}}{1-z_{1}}\right)^ {\frac{5}{2}}-
\frac{1}{s_{2}^{\frac{1}{2}}(1-s_{2})^{2}}\left(\frac{z_{2}}{1-z_{2}}%
\right)^{\frac{5}{2}} \right]\times \\
&\int_{\mathbb{R}^{3}}\frac{f(p^{2})^{2}d\bar{p}}{p^{0}(s_{1}, z_{1})}+
\frac{32\pi}{s_{2}^{\frac{1}{2}}(1-s_{2})^{2}}\left(\frac{z_{2}}{1-z_{2}}%
\right)^{\frac{5}{2}} \int_{\mathbb{R}^{3}}f(p^{2})^{2}\left(\frac{1}{%
p^{0}(s_{1}, z_{1})}-\frac{1}{p^{0}(s_{2}, z_{2})}\right)d\bar{p}
\end{align*}
Notice that:
\[
p^{0}(s_{1}, z_{1})>\left(\frac{2z_{1}}{(1-s_{1})(1-z_{1})}\right)^{\frac{1}{%
2}}|p^{2}| \tag{k}
\]
(4.59) for $P_{2}$ then follows from (i), (k), $|p^{2}|<\sqrt{1+|\bar{p}|^{2}%
}$, (f) and (4.46).

Next, (4.60) follows from the expressions of $P_{1}$, $P_{2}$ in (4.30),
using (j) for the integral in $P_{1}$ and (k) for the integral in $P_{2}$.

Finally to obtain (4.61), we write, for i=1, 2:
\[
\frac{P_{i}(s_{1}, z_{1}, f)}{H_{1}}-\frac{P_{i}(s_{2}, z_{2}, f)}{H_{2}}%
=\left(\frac{1}{H_{1}}-\frac{1}{H_{2}}\right)P_{i}(s_{1}, z_{1}, f)+\frac{1}{%
H_{2}}(P_{i}(s_{1}, z_{1}, f)-P_{i}(s_{2}, z_{2}, f))
\]
and (4.61) is a direct consequence of $\frac{1}{H_{1}}-\frac{1}{H_{2}}=\frac{%
H_{2}-H_{1}}{H_{1}H_{2}}$, (4.59), (4.60) and $0<\alpha(s)<1$ for $0<s<1$.
This completes the proof of lemma \ref{l:4.10}$\blacksquare$\newline

Now, concerning the other terms in (4.57), a usual calculation gives:
\[
\begin{cases}
|(1+\Sigma_{+1}^{2})H_{1}^{2}-(1+\Sigma_{+2}^{2})H_{2}^{2}|\leq
CH_{0}(H_{0}+1)(|H_{1}-H_{2}|+|\Sigma_{+1}-\Sigma_{+2}|)\\
|s_{1}(1-s_{1})\Sigma_{+1}H_{1} - s_{2}(1-s_{2})\Sigma_{+2}H_{2}|\leq
C(|H_{1}-H_{2}|+|s_{1}-s_{2}|+|\Sigma_{+1}-\Sigma_{+2}|)\\
|z_{1}(1-z_{1})(1+\Sigma_{+1}-3s_{1}\Sigma_{+1})H_{1} -
z_{2}(1-z_{2})(1+\Sigma_{+2}-3s_{2}\Sigma_{+2})H_{2}| \leq \\
C(|H_{1}-H_{2}|+|s_{1}-s_{2}|+|z_{1}-z_{2}|+|\Sigma_{+1}-\Sigma_{+2}|)\\
|(1-\Sigma_{+1}^{2})H_{1}\Sigma_{+1}-(1-\Sigma_{+2}^{2})H_{2}\Sigma_{+2}|%
\leq C(1+H_{0})(|H_{1}-H_{2}|+|\Sigma_{+1}-\Sigma_{+2}|)\\ \forall s_{1},
s_{2}, z_{1}, z_{2}\in ]0, 1[; H_{1}, H_{2}\in ]0, H_{0}]; \Sigma_{+1},
\Sigma_{+2}\in ]-1, \frac{1}{2}]. \end{cases}  \tag{4.62}
\]
We then deduce from (4.59), (4.61) and (4.62), using the expression (4.57)
of $F$ the inequality:
\[
\| F(t, H_{1}, s_{1}, z_{1}, \Sigma_{+1})-F(t, H_{2}, s_{2}, z_{2},
\Sigma_{+2}) \| \leq
M(|H_{1}-H_{2}|+|s_{1}-s_{2}|+|z_{1}-z_{2}|+|\Sigma_{+1}-\Sigma_{+2}|) %
\tag{4.63}
\]
where:
\[
M=M(H_{1}, H_{2}, s_{1}, s_{2}, z_{1}, z_{2})=C\left[1+\left(1+\frac{1}{%
H_{2}(1-H_{1})}\right)\frac{\|\ f(t) \|}{\alpha^{6}(s_{1})
\alpha^{5}(s_{2})\alpha^{8}(z_{1})\alpha^{5}(z_{2})}\right]
\tag{4.64}
\]
We can now state:

\begin{proposition}
\label{p:4.11} Let $f \in C([0, T]; X_{r})$, $T>0$, be given. Then, the
initial value problem for the system (IV.2), with initial data  $%
(H_{0}, s_{0}, z_{0}, \Sigma_{+0})$ at $t=0$ satisfying (4.56) has a unique solution $%
( H, s, z, \Sigma_{+})$ on $[0, T]$.
\end{proposition}

\textbf{Proof:} Take, following the definition (4.58) of the domain of $F$:
\[
(H^{0}, s^{0}, z^{0}) \in ]0, H_{0}]\times]0, 1[\times]0, 1[ \tag{4.65}
\]
and:
\[
H_{i} \in ]\frac{H^{0}}{2}, H_{0}]; \, s_{i}\in ]\frac{s^{0}}{2}, \frac{%
s^{0}+1}{2}[; \, z_{i}\in ]\frac{z^{0}}{2}, \frac{z^{0}+1}{2}[; \, i=1, 2 %
\tag{4.66}
\]
Then:
\[
H_{i}>\frac{H^{0}}{2}; \, \frac{s^{0}}{2}<s_{i}<\frac{s^{0}+1}{2}; \, \frac{%
z^{0}}{2}<z_{i}<\frac{z^{0}+1}{2}; \, i=1, 2
\]
and:
\[
\frac{1}{H_{i}}<\frac{2}{H^{0}}; \, \frac{1}{\alpha(s_{i})}<\frac{1}{%
\alpha(s^{0})}; \,\frac{1}{\alpha(z_{i})}<\frac{1}{\alpha(z^{0})}; \, i=1, 2 %
\tag{4.67}
\]
Consequently, the number $M$ defined by (4.64) is bounded on the
neighborhood $]\frac{H^{0}}{2}, H_{0}]\times ]\frac{s^{0}}{2}, \frac{s^{0}+1%
}{2}[\times ]\frac{z^{0}}{2}, \frac{z^{0}+1}{2}[$ of $(H^{0}, s^{0}, z^{0})$
in $\mathbb{R}^{3}$, by a number $M^{0}$, depending only on $H^{0}$, $s^{0}$%
, $z^{0}$, $r$, since $\| f(t) \| <r$, $\forall t \in [0, T]$; the
inequality (4.63) then shows that $F$ is locally Lipschitzian with respect
to the norm of $\mathbb{R}^{4}$. By the standard existence theorem for the
first order differential system, the initial values problem for system
(IV.2) has a unique local solution $(H, s, z, \Sigma_{+})$. Now proposition
\ref{p:4.6} in which we take $t_{0}=0$, $t \in [0, T]$, shows, using
definition (4.46) of $\alpha(s)$, $0<s<1$, that the functions $\frac{1}{H}$,
$\frac{1}{s}$, $\frac{1}{1-s}$, $\frac{1}{z}$, $\frac{1}{1-z}$ are uniformly
bounded and (4.60) implies that the functions $P_{1}$, $P_{2}$ are uniformly
bounded; so $F$ defined by (4.57) is uniformly bounded and, by the standard
theory on the first order differential system, the solution $(H, s, z,
\Sigma_{+})$ is global on $[0, T]$. This ends the proof of prop.\ref{p:4.11}
$\blacksquare$

It shows useful to deduce the following result:

\begin{proposition}
\label{p:4.12} Let $f \in C([0,T]; X_{r})$, $T>0$, be given. Let $(\breve{H},%
\breve{s},\breve{z},\breve{\Sigma}_{+})$ be the solution of the
initial value problem for the system (IV.2) with the initial data
$(H_{0}, s_{0}, z_{0}, \Sigma_{+0})$ at $t=0$, satisfying (4.56).
Let $t_{0}\in [0,T[$. Then
the initial value problem for the system (IV.2), with initial data $(\breve{%
H},\breve{s},\breve{z},\breve{\Sigma}_{+})(t_{0})$ at $t=t_{0}$, has a
unique solution $(H, s, z, \Sigma_{+})$ on $[t_{0}, t_{0}+ \delta]$ where $%
\delta >0$ is independent of $t_{0}$.
\end{proposition}

\textbf{Proof:} The proof of the existence of a local solution
$(H, s, z, \Sigma_{+})$ on an interval $[t_{0}, t_{0}+ \delta]$,
for the initial value
problem for the system (IV.2) with initial data $(\breve{H},\breve{s},\breve{%
z},\breve{\Sigma}_{+})(t_{0})$ is analogous to that of
prop.\ref{p:4.11}. Suppose we look for $\delta$ such that
$0<\delta <1$; then applying the inequalities of prop.\ref{p:4.7}
 leads
to an inequality (4.63) with a constant $M$ independent of the
initial data at $t=t_{0}$, and of $t_{0}$; from there, the
existence of such a number $\delta >0$ $\blacksquare$

We end this section by the following result:

\begin{theorem}
\label{t:4.13} Let $f \in C([0,T]; X_{r})$, $T>0$, be given. Then
the initial value problem for the Einstein equations
(4.1)-(4.2)-(4.3) with initial data\\ \mbox{$(a_{0}, b_{0},
\dot{a}_{0}, \dot{b}_{0}, f_{0})$ at $t=0$},
satisfying (4.13), and the initial constraint (4.12), has a unique solution $%
a$, $b$ on $[0, T]$.
\end{theorem}

\textbf{Proof:} Apply proposition \ref{p:4.11}, choosing by virtue of the
change of variables (4.22), (4.23), (4.24), (4.25) and using (4.16), the
initial data $H_{0}$, $s_{0}$, $z_{0}$, $\Sigma_{+0}$, as follows:
\[
\begin{cases} H_{0}=\frac{1}{3}\left(
\frac{\dot{a}_{0}}{a_{0}}+2\frac{\dot{b}_{0}}{b_{0}} \right); \quad
s_{0}=\frac{b_{0}^{2}}{b_{0}^{2}+2a_{0}^{2}} \\
z_{0}=\frac{1}{a_{0}^{-2}+2b_{0}^{-2}+1}; \quad
\Sigma_{+0}=\frac{1}{H_{0}}\frac{\dot{b}_{0}}{b_{0}}-1 \end{cases}
\tag{4.68}
\]
Notice that $\frac{\dot{a}_{0}}{a_{0}}>0$ $\Rightarrow$ $\frac{\dot{b}_{0}}{%
b_{0}}<\frac{3}{2}H_{0}$; then $-1<\Sigma_{+0}=\frac{1}{H_{0}}\frac{\dot{b}%
_{0}}{b_{0}}-1<\frac{1}{2}$; so that $(H_{0}, s_{0}, z_{0}, \Sigma_{+0})$
defined by (4.68) satisfy the assumption (4.56). Proposition \ref{p:4.11}
then proves the existence of the unique solution $(H, s, z, \Sigma_{+})$ on $%
[0, T]$, of the initial value problem for the system (IV.2) with
the above initial data $(H_{0}, s_{0}, z_{0}, \Sigma_{+0})$ at
$t=0$. The equivalence of the Einstein evolution equations
(4.37)-(4.38) with the system (IV.2), then shows that, $(a, b)$
defined in (4.28) in lemma \ref{l:4.2} is the unique solution on
$[0, T]$ of the initial value problem for the Einstein evolution
equations. Now, since the initial constraint (4.12) is satisfied,
this implies that the Hamiltonian constraint (4.1) is satisfied on
$[0, T]$. This ends the proof of theorem \ref{t:4.13}
$\blacksquare$

\section{Local Existence theorem for the Coupled Einstein-Boltzmann System}

As we proved in paragraph 3 and paragraph 4, the Einstein-Boltzmann System
in $(a, b, f)$ is equivalent to the following first order differential
system in $(f, H, s, z, \Sigma_{+})$ given by (3.15) and (IV.2):

\[
\,\,\,\,\text{(V)}\,\,\,\,\,\,\,\,\,\,\,\,\,\,\,\,\,\,\left\{
\begin{array}{l}
\frac{df}{\,dt}=\frac{1}{p^{o}}\,Q(f,f)\,\,\,\,\,\,\,\,\,\,\,\,\,\,\,\,\,\,%
\,\,\,\,\,\,\,\,\,\,\,\,\,\,\,\,\,\,\,\,\,\,\,\,\,\,\,\,\,\,\,\,\,\,\,\,\,\,%
\,\,\,\,\,\,\,\,\,\,\,\,\,\,\,\,\,\,\,\,\,\,\,\,\,\,\,\,\,\,\,\,\,\,\,\,\,\,%
\,\,\,\,\,\,\,\,\,\,\,\,\,\,\text{(5.1)} \\
\frac{dH}{dt}=-\frac{3}{2}(1+\Sigma _{+}^{2})H^{2}-\frac{P_{1}+2P_{2}}{6}%
\qquad \qquad \qquad \qquad \qquad \,\,\,\,\,\,\,\text{(5.2)} \\
\frac{ds}{dt}=6s(1-s)\Sigma _{+}H\qquad
\,\,\,\,\,\,\,\,\,\,\,\,\,\,\,\,\,\,\,\,\,\,\,\,\,\,\,\,\,\,\,\,\,\,\,\,\,\,%
\,\,\,\,\,\,\,\,\,\,\,\,\,\,\,\,\,\,\,\,\,\,\,\,\,\,\,\,\,\,\,\,\,\,\,\,\,\,%
\,\,\,\,\,\,\,\,\,\text{(5.3)} \\
\frac{dz}{dt}=2z(1-z)(1+\Sigma _{+}-3s\Sigma _{+})H\qquad \quad \qquad
\qquad \qquad \,\,\,\,\,\,\,\text{(5.4)} \\
\frac{d\Sigma _{+}}{dt}=-\frac{3}{2}(1-\Sigma _{+}^{2})H\Sigma _{+}+\frac{%
P_{1}}{6H}(\Sigma _{+}-2)+\frac{P_{2}}{3H}(\Sigma _{+}+1)\,.\,\,\,\,\,\text{%
(5.5)}
\end{array}
\right.
\]
in which $Q$ is the collision operator defined by (3.1)-(3.2)-(3.3) and $%
p^{0}$, $P_{1}$, $P_{2}$ are defined in terms of $s, z, f$ by (4.30). To
solve system (V), we apply the standard theory of the first order
differential system for functions on $\mathbb{R}$, with values in the Banach
space $E=L^{1}_{2}(\mathbb{R}^{3}) \times\mathbb{R}\times\mathbb{R}\times%
\mathbb{R}\times\mathbb{R}$, whose norm will be taken to be: $\| (f, H, s,
z, \Sigma_{+}) \|_{E}=\| f \|+|H|+|s|+|z|+|\Sigma_{+}|$. Since $f$ is now
also an unknown, the problem will be to prove that the function $\tilde{F}$,
defined by the r.h.s of (V) i.e:
\begin{equation}
\tilde{F}(f,H,\,s,\,z,\,\Sigma _{+})=\left\{
\begin{array}{l}
\frac{1}{p^{o}}\,Q(f,f)\, \\
-\frac{3}{2}(1+\Sigma _{+}^{2})H^{2}-\frac{P_{1}+2P_{2}}{6} \\
6s(1-s)\Sigma _{+}H\qquad \qquad \\
2z(1-z)(1+\Sigma _{+}-3s\Sigma _{+})H \\
-\frac{3}{2}(1-\Sigma _{+}^{2})H\Sigma _{+}+\frac{P_{1}}{6H}(\Sigma _{+}-2)+%
\frac{P_{2}}{3H}(\Sigma _{+}+1)
\end{array}
\right.  \tag*{(5.6)}
\end{equation}
which does not depend explicitly on $t$, is locally Lipschitzian in $(f, H,
s, z, \Sigma_{+})$, with respect to the above norm of $E$. Following (4.58),
$\tilde{F}$ will be defined on: $L^{1}_{2}(\mathbb{R}^{3})\times ]0, H_{0}]
\times ]0, 1[ \times ]0, 1[ \times ]-1, \frac{1}{2}]$. A glance to (5.6)
shows that, taking into account the study we did in paragraph 4 for the
system (IV.2), the new problem to face here will be the study of all the
differences in $f$, and the differences of $\frac{Q(f,f)}{p^{0}}$ in $s$ and
$z$, which turns out to be the only real problem we meet now, since
expressions $P_{1}$, $P_{2}$ in (4.30) show that these functions are linear
in $f$. We then begin by proving:

\begin{lemma}
\label{l:5.1} Let $f_{1}, f_{2} \in L^{1}_{2}(\mathbb{R}^{3})$, $s_{1},
s_{2}, z_{1}, z_{2} \in ]0, 1[$ then:
\[
\left\| \left( \frac{1}{p^{0}(s_{1},z_{1})}-\frac{1}{p^{0}(s_{2},z_{2})}
\right)Q(f_{1},f_{1})(s_{2},z_{2}) \right\| \leq \frac{C\|f_{1}%
\|^{2}(|s_{1}-s_{2}|+|z_{1}-z_{2}|)}{\alpha^{4}(s_{1})
\alpha^{2}(s_{2})\alpha^{2}(z_{1})\alpha^{3}(z_{2})} \tag{5.7}
\]
\[
\left\|\frac{1}{p^{0}(s_{2},z_{2})}%
(Q(f_{1},f_{1})(s_{2},z_{2})-Q(f_{2},f_{2})(s_{2},z_{2})) \right\| \leq
\frac{C(\|f_{1}\|+\|f_{2}\|)\|f_{1}-f_{2}\|}{ \alpha^{2}(s_{2})%
\alpha^{2}(z_{2})} \tag{5.8}
\]
where $C>0$ is a constant.
\end{lemma}

\textbf{Proof:} For (5.7), we follow the proof of (4.59) in lemma \ref
{l:4.10}, using this time in that proof $(p^{0}(s_{1}, z_{1}))^{2}>\frac{z_{1}}{%
s_{1}(1-z_{1})}|p^{1}|^{2}$ and $(p^{0}(s_{1},z_{1}))^{2}>\frac{2z_{1}}{%
(1-s_{1})(1-z_{1})}[(p^{2})^{2}+(p^{3})^{2}]$ and, the relations
(a) and (d) to obtain:
\[
\left|\frac{1}{p^{0}(s_{1},z_{1})}- \frac{1}{p^{0}(s_{2},z_{2})}
\right| \leq
\frac{C(|s_{1}-s_{2}|+|z_{1}-z_{2}|)}{p^{0}(s_{2},z_{2})z_{1}(1-z_{1})(1-z_{2})}
\left( \frac{1}{s_{1}s_{2}} +\frac{1}{(1-s_{1})(1-s_{2})} \right) %
\tag{a}
\]
Now, we use (3.27) with $f=f_{1}$, $g=0$, (3.28) and (4.28) which gives the
expression of $a$, $b$ in terms of $s$ and $z$ to obtain:
\[
\left\| \frac{Q(f_{1}, f_{1})(s_{2}, z_{2})}{p^{0}(s_{2},z_{2})} \right\|
\leq \frac{C \| f_{1} \|^{2}}{s_{2}^{\frac{1}{2}}(1-s_{2})(1-z_{2})^{\frac{3%
}{2}}} \tag{b}
\]
(5.7) then follows from (a) and (b) and definition (4.46) of $\alpha(s)$.
Now (5.8) is given directly by (3.27) with $f=f_{1}$, $g=f_{2}$, the
expression of $ab^{2}$ in terms of $s$, $z$, and once more (4.46). This
completes the proof of lemma \ref{l:5.1} $\blacksquare$ \newline
We will also need the following result which is not obvious:

\begin{lemma}
\label{l:5.2} Let $f \in L^{1}_{2}(\mathbb{R}^{3})$, $s_{1}$, $s_{2}$, $%
z_{1} $, $z_{2}$ $\in ]0, 1[$; then:
\[
\left\|\frac{1}{p^{0}(s_{1},z_{1})}(Q(f,f)(s_{1},z_{1})-Q(f,f)(s_{2},z_{2}))
\right\| \leq \frac{C\|f\|^{2}(|s_{1}-s_{2}|+|z_{1}-z_{2}|)}{%
\alpha^{3}(s_{1}) \alpha^{3}(s_{2})\alpha^{5}(z_{1})\alpha^{3}(z_{2})} %
\tag{5.9}
\]
where $C>0$ is a constant.
\end{lemma}

\textbf{Proof:} Here, we split $Q$ into $Q^{+}$, $Q^{-}$ using the basic
definition (3.1) of $Q$, i.e $Q=Q^{+}-Q^{-}$ where $Q^{+}$, $Q^{-}$ are
defined by (3.2) and (3.3) in which $ab^{2}$ is expressed in terms of $s$
and $z$. It is here that we need the Lipschitz continuity assumptions (3.7),
(3.8) on the collision kernel $A$. We write:
\begin{align*}
Q(f,f)(s_{1},z_{1})-Q(f,f)(s_{2},z_{2})=[Q^{+}(f,f)(s_{1},z_{1})-Q^{+}(f,f)(s_{2},z_{2})]+
\\
[Q^{-}(f,f)(s_{2},z_{2})-Q^{-}(f,f)(s_{1},z_{1})]  \tag{a}
\end{align*}
We can write, using the expression (3.2) of $Q^{+}$:
\[
Q^{+}(f,f)(s_{1},z_{1})-Q^{+}(f,f)(s_{2},z_{2})= \int_{\mathbb{R}^{3}\times
S^{2}}\Omega(s_{1},s_{2},z_{1},z_{2},\bar{p},\bar{q},\bar{p^{\prime}},\bar{%
q^{\prime}})f(\bar{p^{\prime}})f(\bar{q^{\prime}}) d\bar{q}d\omega \tag{b}
\]
where $\Omega$ is a function of the indicated arguments, we write on the
form:
\[
\Omega = \left( \frac{ab^{2}}{q^{0}}
\right)(s_{1},z_{1})[A(s_{1},z_{1})-A(s_{2},z_{2})]+\left[ \left(\frac{ab^{2}%
}{q^{0}}\right)(s_{1},z_{1})-\left(\frac{ab^{2}}{q^{0}}\right)(s_{2},z_{2})
\right]A(s_{2},z_{2}) \tag{c}
\]
in which $A(s_{i},z_{i})$ stands in fact for $A(a(s_{i},z_{i}),
b(s_{i},z_{i}), \bar{p},\bar{q},\bar{p^{\prime}},\bar{q^{\prime}})$ i=1, 2;
we now bound the first term in (c). The assumptions (3.7), (3.8) on the
collision kernel $A$ give, adding and subtracting $A(a(s_{2},z_{2}),
b(s_{1},z_{1}))$:
\[
|A(s_{1},z_{1})-A(s_{2},z_{2})| \leq
k_{0}(|a(s_{1},z_{1})-a(s_{2},z_{2})|+|b(s_{1},z_{1})-b(s_{2},z_{2})|) %
\tag{d}
\]
Now (4.28) gives, by usual calculation:
\[
\begin{cases}
|a(s_{1},z_{1})-a(s_{2},z_{2})|=\left|\sqrt{\frac{z_{1}}{s_{1}(1-z_{1})}}-
\sqrt{\frac{z_{2}}{s_{2}(1-z_{2})}} \right| \leq
\frac{C(|s_{1}-s_{2}|+|z_{1}-z_{2}|)}{s_{1}s_{2}z_{1}^{%
\frac{1}{2}}(1-z_{1})(1-z_{2})}\\
|b(s_{1},z_{1})-b(s_{2},z_{2})|=\left|\sqrt{%
\frac{2z_{1}}{(1-s_{1})(1-z_{1})}}- \sqrt{\frac{2z_{2}}{(1-s_{2})(1-z_{2})}}
\right| \leq
\frac{C(|s_{1}-s_{2}|+|z_{1}-z_{2}|)}{(1-s_{1})(1-s_{2})z_{1}^{%
\frac{1}{2}}(1-z_{1})(1-z_{2})} \end{cases}  \tag{e}
\]
We then deduce from (d), (e) and (4.28) which gives:
\[
(ab^{2})(s,z)=\frac{2}{s^{\frac{1}{2}}}\left(\frac{z}{1-z}\right)^{\frac{3}{2%
}} \tag{f}
\]
that:
\[
\left| \left( \frac{ab^{2}}{q^{0}}
\right)(s_{1},z_{1})[A(s_{1},z_{1})-A(s_{2},z_{2})] \right| \leq \frac{%
C(|s_{1}-s_{2}|+|z_{1}-z_{2}|)}{\alpha^{3}(s_{1})\alpha^{2}(s_{2})%
\alpha^{3}(z_{1}) \alpha(z_{2})q^{0}(s_{1}, z_{1})} \tag{g}
\]
Now, we bound the second term in (c); we write:
\begin{align*}
\left(\frac{ab^{2}}{q^{0}}\right)(s_{1},z_{1})-\left(\frac{ab^{2}}{q^{0}}%
\right)(s_{2},z_{2}) =\frac{%
(ab^{2})(s_{1},z_{1})q^{0}(s_{2},z_{2})-(ab^{2})(s_{2},z_{2})q^{0}(s_{1},z_{1})%
} {q^{0}(s_{1},z_{1})q^{0}(s_{2},z_{2})} \\
=\frac{1}{q^{0}(s_{1},z_{1})}\left[(ab^{2})(s_{1},z_{1}) - \frac{%
q^{0}(s_{1},z_{1})}{q^{0}(s_{2},z_{2})}(ab^{2})(s_{2},z_{2})\right] \\
=\frac{1}{q^{0}(s_{1},z_{1})}\left[(ab^{2})(s_{1},z_{1})
-(ab^{2})(s_{2},z_{2})+ \left(1- \frac{q^{0}(s_{1},z_{1})}{q^{0}(s_{2},z_{2})%
}\right)(ab^{2})(s_{2},z_{2})\right]
\end{align*}
Hence:
\begin{align*}
\left( \frac{ab^{2}}{q^{0}} \right)(s_{1},z_{1})-\left( \frac{ab^{2}}{q^{0}}
\right)&(s_{2},z_{2}) =\frac{1}{q^{0}(s_{1},z_{1})}\times \\
&\left[(ab^{2})(s_{1},z_{1})-(ab^{2})(s_{2},z_{2})+ \left( \frac{%
q^{0}(s_{2},z_{2})-q^{0}(s_{1},z_{1})}{q^{0}(s_{2},z_{2})}\right)(ab^{2})
(s_{2},z_{2}) \right]  \tag{h}
\end{align*}
Now, by (4.30):
\begin{align*}
\frac{q^{0}(s_{2},z_{2})-q^{0}(s_{1},z_{1})}{q^{0}(s_{2},z_{2})}&=\frac{%
\left( \frac{z_{2}}{s_{2}(1-z_{2})}-\frac{z_{1}}{s_{1}(1-z_{1})}%
\right)(q^{1})^{2}} {q^{0}(s_{2},z_{2})%
\left[q^{0}(s_{1},z_{1})+q^{0}(s_{2},z_{2})\right]} \\
&+ \frac{\left( \frac{2z_{2}}{(1-s_{2})(1-z_{2})}-\frac{2z_{1}}{%
(1-s_{1})(1-z_{1})}\right)\left[(q^{2})^{2} +(q^{3})^{2}\right]}{%
q^{0}(s_{2},z_{2})\left[q^{0}(s_{1},z_{1})+q^{0}(s_{2},z_{2})\right]}
\end{align*}
and we proceed as in the proof of (4.59), using this time:
\[
\left[q^{0}(s_{2},z_{2})\right]^{2}\geq\frac{z_{2}}{s_{2}(1-z_{2})}%
(q^{1})^{2};\quad \left[q^{0}(s_{2},z_{2})\right]^{2}\geq\frac{2z_{2}}{%
(1-s_{2})(1-z_{2})}\left[(q^{2})^{2} +(q^{3})^{2}\right]
\]
to obtain, using expression (f) of $ab^{2}$:
\[
(ab^{2})(s_{2},z_{2})\left| \frac{q^{0}(s_{2},z_{2})-q^{0}(s_{1},z_{1})}{%
q^{0}(s_{2},z_{2})} \right| \leq \frac{C(|s_{1}-s_{2}|+|z_{1}-z_{2}|)}{%
\alpha^{2}(s_{1})\alpha^{3}(s_{2})\alpha^{2}(z_{1}) \alpha^{3}(z_{2})} %
\tag{i}
\]
Now, concerning the first term in (h), we have, using (f) and by usual
calculation:
\begin{align*}
|(ab^{2})(s_{1},z_{1})-(ab^{2})(s_{2},z_{2})|&=2\left| \frac{1}{s_{1}^{\frac{%
1}{2}}\left( \frac{z_{1}}{1-z_{1}} \right)^{\frac{3}{2}}}-\frac{1}{s_{2}^{%
\frac{1}{2}}\left( \frac{z_{2}}{1-z_{2}} \right)^{\frac{3}{2}}} \right| \\
& \leq \frac{C(|s_{1}-s_{2}|+|z_{1}-z_{2}|)}{\alpha(s_{1})\alpha(s_{2})%
\alpha^{5}(z_{1}) \alpha^{3}(z_{2})}  \tag{j}
\end{align*}
We then have for the second term in (c), using (h), (i), (j) and $0 \leq A
\leq C_{0}$:
\[
A(s_{2},z_{2})\left| \left(\frac{ab^{2}}{q^{0}}\right)(s_{1},z_{1})-\left(%
\frac{ab^{2}}{q^{0}}\right)(s_{2},z_{2}) \right| \leq \frac{%
C(|s_{1}-s_{2}|+|z_{1}-z_{2}|)}{q^{0}(s_{1},z_{1})\alpha^{2}(s_{1})%
\alpha^{3}(s_{2}) \alpha^{5}(z_{1})\alpha^{3}(z_{2})} \tag{k}
\]
We then deduce from (b), using (c), (g) and (k) that:
\begin{align*}
&\left\|\frac{1}{p^{0}(s_{1},z_{1})}%
(Q^{+}(f,f)(s_{1},z_{1})-Q^{+}(f,f)(s_{2},z_{2})) \right\| \leq \\
&\frac{C(|s_{1}-s_{2}|+|z_{1}-z_{2}|)}{\alpha^{3}(s_{1})
\alpha^{3}(s_{2})\alpha^{5}(z_{1})\alpha^{3}(z_{2})}\int_{\mathbb{R}^{3}}
\int_{\mathbb{R}^{3}}\int_{S^{2}}\frac{\sqrt{1+|\bar{p}|^{2}}\left|f(\bar{%
p^{\prime}})\right|
\left|f(\bar{q^{\prime}})\right|}{p^{0}q^{0}(s_{1},z_{1})}d\bar{p}d\bar{q}d\omega
\tag{l}
\end{align*}
We compute this integral the same way as in the proof of (3.25) in lemma \ref
{l:3.4}, using the change of variables $(\bar{p}, \bar{q})\, \mapsto (\bar{%
p^{\prime}}, \bar{q^{\prime}})$ defined by (3.12) to obtain:
\[
\left\|\frac{1}{p^{0}(s_{1},z_{1})}%
(Q^{+}(f,f)(s_{1},z_{1})-Q^{+}(f,f)(s_{2},z_{2})) \right\| \leq \frac{%
C\|f\|^{2}(|s_{1}-s_{2}|+|z_{1}-z_{2}|)}{\alpha^{3}(s_{1})
\alpha^{3}(s_{2})\alpha^{5}(z_{1})\alpha^{3}(z_{2})} \tag{m}
\]
We now proceed the same way for the second term in (a), i.e\\
\mbox{$ (Q^{-}(f,f)(s_{2},z_{2})-Q^{-}(f,f)(s_{1},z_{1}))$}; the
only difference is that, in the integrals (b) and (l),
$f(\bar{p^{\prime}})f(\bar{q^{\prime}})$ is replaced by
$f(\bar{p})f(\bar{q})$. So, no need this time to change the
variables and a direct calculation using $p^{0}q^{0} \geq 1$ leads
to the same estimation (m), just substituting $Q^{-}$ to $Q^{+}$,
and lemma \ref{l:5.2} follows $\blacksquare$

Now we deduce from lemma \ref{l:5.1} and lemma \ref{l:5.2}, the following:

\begin{lemma}
\label{l:5.3} Let $f_{1}$, $f_{2} \in L^{1}_{2}(\mathbb{R}^{3})$, $s_{1}$, $%
s_{2}$, $z_{1}$, $z_{2}$ $\in ]0, 1[$; then:
\begin{align*}
\left\|\frac{Q(f_{1},f_{1})(s_{1},z_{1})}{p^{0}(s_{1},z_{1})}-\frac{%
Q(f_{2},f_{2}) (s_{2},z_{2})}{p^{0}(s_{2},z_{2})} \right\|& \leq \frac{%
C\left(\|f_{1}\|^{2} +\|f_{1}\|+\|f_{2}\|
\right)}{\alpha^{4}(s_{1})
\alpha^{3}(s_{2})\alpha^{5}(z_{1})\alpha^{3}(z_{2})} \times \\
&(\|f_{1}-f_{2}\| +|s_{1}-s_{2}|+|z_{1}-z_{2}|)  \tag{5.10}
\end{align*}
where $C>0$ is a constant.
\end{lemma}

\textbf{Proof:} Write:
\begin{align*}
\frac{Q(f_{1},f_{1})(s_{1},z_{1})}{p^{0}(s_{1},z_{1})}-\frac{Q(f_{2},f_{2})
(s_{2},z_{2})}{p^{0}(s_{2},z_{2})}=\frac{Q(f_{1},f_{1})(s_{1},z_{1})-
Q(f_{1},f_{1})(s_{2},z_{2})}{p^{0}(s_{1},z_{1})} \\
+\left(\frac{1}{p^{0}(s_{1},z_{1})}-\frac{1}{p^{0}(s_{2},z_{2})}\right)
Q(f_{1},f_{1})(s_{2},z_{2}) \\
+\frac{Q(f_{1},f_{1})(s_{2},z_{2})- Q(f_{2},f_{2})(s_{2},z_{2})}{%
p^{0}(s_{2},z_{2})}
\end{align*}
and apply: (5.9) with $f=f_{1}$ to the first term, (5.7) to the second term,
and (5.8) to the third term and $0<\alpha(s)<1$, to obtain (5.10) $%
\blacksquare$

Finally, concerning the differences in $P_{i}$ and $\frac{P_{i}}{H}$, we
prove:

\begin{lemma}
\label{l:5.3} Let $f_{1}$, $f_{2} \in L^{1}_{2}(\mathbb{R}^{3})$; $s_{1}$, $%
s_{2}$, $z_{1}$, $z_{2}$ $\in ]0, 1[$; $H_{1}$, $H_{2} \in ]0, H_{0}]$.
Then:
\[
\left| P_{i}(s_{1},z_{1},f_{1})-P_{i}(s_{2},z_{2},f_{2}) \right| \leq \frac{%
C\left(1+\|f_{2}\| \right)\left( \|f_{1}-f_{2}\|
+|s_{1}-s_{2}|+|z_{1}-z_{2}| \right)}{\alpha^{6}(s_{1})
\alpha^{5}(s_{2})\alpha^{8}(z_{1})\alpha^{5}(z_{2})} \tag{5.11}
\]
\[
\left| \frac{P_{i}(s_{1},z_{1},f_{1})}{H_{1}}-\frac{P_{i}(s_{2},z_{2},f_{2})%
}{H_{2}} \right| \leq \frac{C\left(1+\|f_{1}\| \right)\left( \|f_{1}-f_{2}\|
+|s_{1}-s_{2}|+|z_{1}-z_{2}|+ |H_{1}-H_{2}| \right)}{ H_{2}(1+H_{1})
\alpha^{6}(s_{1}) \alpha^{5}(s_{2})\alpha^{8}(z_{1})\alpha^{5}(z_{2})} %
\tag{5.12}
\]
i=1, 2; where $C>0$ is a constant.
\end{lemma}

\textbf{Proof:} The expressions of $P_{i}$, i=1, 2 in (4.30) show that $%
P_{i} $ is linear in $f$. So we can write:
\[
P_{i}(s_{1},z_{1},f_{1})-P_{i}(s_{2},z_{2},f_{2})=
P_{i}(s_{1},z_{1},f_{1}-f_{2}) + \left( P_{i}(s_{1},z_{1},f_{2})-
P_{i}(s_{2},z_{2},f_{2}) \right) \tag{a}
\]
\[
\frac{P_{i}(s_{1},z_{1},f_{1})}{H_{1}}- \frac{P_{i}(s_{2},z_{2},f_{2})}{H_{2}%
}= + \left( \frac{P_{i}(s_{1},z_{1},f_{1})}{H_{1}}- \frac{%
P_{i}(s_{2},z_{2},f_{1})}{H_{2}} \right)+\frac{P_{i}(s_{2},z_{2},f_{1}-f_{2})%
}{H_{2}} \tag{b}
\]
so, to obtain (5.11), apply (4.60) with $f(t)=f_{1}-f_{2}$ to the first term
of (a), and (4.59) with $f(t)=f_{2}$ to the second term of (a). To obtain
(5.12), apply (4.61) with $f(t)=f_{1}$ to obtain the first term of (b), and
(4.60) with $f(t)=f_{1}-f_{2}$ to the second term of (b) $\blacksquare$

We can then write, using (4.62), (5.10), (5.11), (5.12) and the expression
(5.6) of $\tilde{F}$:
\begin{align*}
& \| \tilde{F}(f_{1},H_{1},s_{1},z_{1},\Sigma_{+1})-\tilde{F}%
(f_{2},H_{2},s_{2}, z_{2},\Sigma_{+2}) \|_{E} \leq \\
& N \left( \|f_{1}-f_{2}\| +|s_{1}-s_{2}|+|z_{1}-z_{2}|+ |H_{1}-H_{2}| +
|\Sigma_{+1}-\Sigma_{+2}| \right)  \tag{5.13}
\end{align*}
where:
\begin{align*}
&N=N(f_{1},f_{2},H_{1},H_{2},s_{1},s_{2},z_{1},z_{2}) \\
&=C\left[1+\left(1+\frac{1}{H_{2}(1+H_{1})} \right) \left( \frac{%
\left(1+\|f_{1}\|+\|f_{2}\|+\|f_{1}\|^{2} \right)}{\alpha^{6}(s_{1})
\alpha^{5}(s_{2})\alpha^{8}(z_{1})\alpha^{5}(z_{2})} \right) \right]
\tag{5.14}
\end{align*}
We can now state:

\begin{proposition}
\label{p:5.5} There exists a number $l>0$, such that, the initial
value problem for the differential system (V), with initial data
$(f_{0}, H_{0}, s_{0}, z_{0}, \Sigma_{+0})$ at $t=0$, satisfying
(4.56) and $f_{0} \in L_{2}^{1}(\mathbb{R}^{3})$, has a unique
solution $(f, H, s, z, \Sigma_{+})$ on $[0, l]$.
\end{proposition}

\textbf{Proof:} Choose $H^{0}$, $s^{0}$, $z^{0}$ as in (4.65) and
$f^{0} \in L_{2}^{1}(\mathbb{R}^{3})$.\newline Take $H_{i}$,
$s_{i}$, $z_{i}$, i=1, 2 satisfying (4.66) and\\ $f_{1}$, $f_{2}
\in B(f^{0},1)=\{ f \in L_{2}^{1}(\mathbb{R}^{3}), \| f-f^{0} \|
\leq 1 \}$. Then $H_{i}$, $s_{i}$, $z_{i}$, i=1, 2 satisfy (4.67)
and $\| f_{i} \| \leq \| f^{0}\|+1 $, i=1, 2. Consequently, the
number $N$ defined by (5.14) is bounded in the neighborhood
$B(f^{0},1) \times ]\frac{H^{0}}{2}, H_{0}]
\times ]\frac{s^{0}}{2}, \frac{s^{0}+1}{2}[ \times ]\frac{z^{0}}{2}, \frac{%
z^{0}+1}{2}[$ of $(f^{0}, H^{0},s^{0},z^{0})$ in $L_{2}^{1}(\mathbb{R}%
^{3})\times \mathbb{R}^{3}$, by a number $N^{0}$, depending only on $f^{0},
H^{0},s^{0},z^{0}$. The inequality (5.13) then shows that $\tilde{F}$ is
locally Lipschitzian with respect to the norm of $E=L_{2}^{1}(\mathbb{R}%
^{3})\times \mathbb{R}^{4}$ and proposition \ref{p:5.5} follows from the
standard existence theorem for the first order differential system, for
functions with values in a Banach space. Notice that $f \in C\left([0, l];
L_{2}^{1}(\mathbb{R}^{3})\right)$ $\blacksquare$

We end this paragraph by the following result:

\begin{theorem}
\label{t:5.6} Let $f_{0} \in L_{2}^{1}(\mathbb{R}^{3})$; $a_{0}$, $b_{0}$, $%
\dot{a}_{0}$, $\dot{b}_{0}$ satisfying (4.13), the initial
constraint (4.12), and $a_{0} \geq \frac{3}{2}$, $b_{0} \geq
\frac{3}{2}$, be given. Let $r> \|f_{0}\|$.\newline Then, there
exists a number $l>0$, such that, the initial value problem for
the coupled Einstein-Boltzmann system (2.13)-(4.1)-(4.2)-(4.3) has
a unique solution $f, a, b$ on $[0,l]$. The solution $(f, a, b)$
has the following properties:
\[
a \quad \text{and} \quad b \quad \text{are increasing functions} %
\tag{5.15}
\]
\[
f \in C\left( [0, l]; X_{r} \right) \tag{5.16}
\]
\[
\|| f \|| \leq \| f_{0} \| \tag{5.17}
\]
\end{theorem}

\textbf{Proof:} Apply proposition \ref{p:5.5}, choosing $f_{0} \in L_{2}^{1}(%
\mathbb{R}^{3})$, $f_{0} \geq 0$ a.e, and define $H_{0}$, $s_{0}$, $z_{0}$, $%
\Sigma_{+0}$ by (4.68); (4.56) is then satisfied. The existence of a unique
solution $(f, a, b)$ of (2.13)-(4.1)-(4.2)-(4.3) on $[0, l]$, $l>0$, is a
direct consequence of proposition \ref{p:5.5}, the equivalence of the above
system and the system (V), and the hypothesis on the initial data at $t=0$.

Now, concerning the properties of the solution $(f, a, b)$:
\begin{itemize}
\item  i) (5.15) is given by corollary \ref{c:4.8} with $t_{0}=0$, $\delta=l$.%

\item  ii) The hypothesis $a_{0} \geq \frac{3}{2}$, $b_{0} \geq
\frac{3}{2}$
and (5.15) show that $a(t) \geq \frac{3}{2}$, $b(t) \geq \frac{3}{2}$, $%
\forall t \in [0, l]$; $a$ and $b$ then satisfy the hypotheses of
the existence theorem \ref{t:3.2}, of the Boltzmann equation in
$f$, in which we take $t_{0}=0$ and \mbox{$T=l>0$.} (5.16) is then
a consequence of the uniqueness and of $C\left( [0, l]; X_{r}
\right) \subset C\left( [0, l]; L^{1}_{2}(\mathbb{R}^{3}) \right)$

\item  iii) (5.17) is given by (3.19) with $t_{0}=0$
$\blacksquare$
\end{itemize}

\section{The Global Existence Theorem for the Coupled Einstein-Boltzmann
System}

\subsection{The Method}

We assume, in all what follows, that the initial data $(f_{0}, a_{0}, b_{0},
\dot{a}_{0}, \dot{b}_{0})$ at $t=0$, for the Einstein-Boltzmann system
(2.13)-(4.1)-(4.2)-(4.3) satisfy (4.12), (4.13), $a_{0} \geq \frac{3}{2}$, $%
b_{0} \geq \frac{3}{2}$, and that, in the initial data $(f_{0}, H_{0},
s_{0}, z_{0}, \Sigma_{+0})$ at $t=0$, for the system (V), paragraph 5, $%
H_{0} $, $s_{0}$, $z_{0}$, $\Sigma_{+0}$ are defined by (4.68),
and hence satisfy (4.56). Denote $[0, T[$, $T>0$, the
\textbf{maximal} existence domain of the solution, denoted here
$(\tilde{f}, \tilde{H}, \tilde{s}, \tilde{z}, \tilde{\Sigma}_{+})$
and given by proposition \ref{p:5.5}, of the initial value problem
for (V), with the above initial data at $t=0$.

If $T= + \infty $, the problem of the global existence is solved. We are
going to show that, if we suppose $T< + \infty $, then the solution $(\tilde{%
f}, \tilde{H}, \tilde{s}, \tilde{z}, \tilde{\Sigma}_{+})$ can be extended
beyond $T$ , which contradicts the maximality of $T$.

The strategy is the following:\newline
Suppose $0<T<+ \infty$ and let $t_{0} \in [0, T[$. We will show that there
exists a strictly positive number $\delta >0$, \textbf{independent of $%
t_{0}$}, such that, the initial value problem for the system (V), on $%
[t_{0}, t_{0}+\delta]$, with initial data \newline
\mbox{$(\tilde{f}(t_{0}), \tilde{H}(t_{0}),
\tilde{s}(t_{0}), \tilde{z}(t_{0}), \tilde{\Sigma}_{+}(t_{0}))$} at $t=t_{0}$%
, has a unique solution $(f, H, s, z, \Sigma_{+})$ on $[t_{0}, t_{0}+\delta]$%
. Then, by taking $t_{0}$ sufficiently close to $T$, for example, $t_{0}$
such that, $0<T-t_{0}<\frac{\delta}{2}$, hence $T<t_{0}+\frac{\delta}{2}$,
we can extend the solution $(\tilde{f}, \tilde{H}, \tilde{s}, \tilde{z},
\tilde{\Sigma}_{+})$ to $[0, t_{0}+\frac{\delta}{2}[$, which contains
strictly $[0, T[$, and this contradicts the maximality of $T$. In order to
simplify the notations, it will be enough to look for a number $\delta$ such
that $0<\delta<1$. As in previous paragraphs, we denote $r$, a real number
such that $r> \|f_{0}\|$.

\subsection{The functional framework}

\begin{proposition}
\label{p:6.1} Let $t_{0} \in [0, T[$ and \, $0<\delta<1$. Then,
any solution $(f, H, s, z, \Sigma_{+})$ for the initial value
problem for the system (V) on $[t_{0}, t_{0}+\delta]$, with the
initial data at $t=t_{0}$:
\[
(f, H, s, z, \Sigma_{+})(t_{0})=(\tilde{f}(t_{0}), \tilde{H}(t_{0}), \tilde{s%
}(t_{0}), \tilde{z}(t_{0}), \tilde{\Sigma}_{+}(t_{0})) \tag{6.1}
\]
satisfy the inequalities:
\[
\frac{1}{H(t_{0}+t)} \leq M_{0};\quad \frac{1}{\alpha(s(t_{0}+t))}\leq
M_{0};\quad \frac{1}{\alpha(z(t_{0}+t))} \leq M_{0}; \quad t \in [0,\delta] %
\tag{6.2}
\]
where:
\[
M_{0}=M_{0}(a_{0},b_{0},\dot{a}_{0},\dot{b}_{0},T)=\left(\frac{1}{H_{0}} +%
\frac{1}{s_{0}} +\frac{1}{z_{0}} \right)e^{10H_{0}(T+1)} \tag{6.3}
\]
in which $H_{0}$, $s_{0}$, $z_{0}$ are defined in terms of $a_{0},b_{0},\dot{%
a}_{0},\dot{b}_{0}$ by (4.68).
\end{proposition}

\textbf{Proof:} Apply proposition \ref{p:4.7} to the subsystem (IV.2) of
(V), and consider the solution $(\breve{H},\breve{s},\breve{z},\breve{\Sigma}%
_{+})=(\tilde{H}, \tilde{s}, \tilde{z}, \tilde{\Sigma}_{+})$ of
the initial value problem (IV.2) on $[0, T[$, with initial data
$(H_{0}, s_{0}, z_{0},
\Sigma_{+0})$, defined by (4.68) in terms $a_{0},b_{0},\dot{a}_{0},\dot{b}%
_{0}$; since $t_{0} \in [0, T[$ and $( H, s, z, \Sigma_{+})(t_{0})=(\tilde{H}%
, \tilde{s}, \tilde{z}, \tilde{\Sigma}_{+})(t_{0})= (\breve{H},\breve{s},%
\breve{z},\breve{\Sigma}_{+})(t_{0})$, (6.2) and (6.3) are given
by (4.52), (4.53), (4.54), (4.55), (4.68), and $0<\delta<1$
$\blacksquare$

In all what follows, $M_{0}$ is the absolute constant defined by (6.3). We
deduce from (4.46), (6.2) and the expressions of $a^{2}$, $b^{2}$ in (4.28)
that:
\[
\frac{1}{s} \leq M_{0}; \quad \frac{1}{z} \leq M_{0}; \quad a^{2}\leq
M_{0}^{2}; \quad b^{2}\leq 2M_{0}^{2} \tag{6.4}
\]
We then deduce from (6.4), the definition (4.23), (4.24) of $z$ and $s$, in
terms of $a$ and $b$, using $a^{2} \geq a_{0}^{2} \geq \frac{9}{4}>2$, and
the inequality for $H$ in (6.2):
\[
\frac{1}{M_{0}} \leq z \leq \frac{1}{1+2M_{0}^{-2}}; \quad \frac{1}{M_{0}}
\leq s =\frac{1}{1+2a^{2}b^{-2}} \leq \frac{1}{1+2M_{0}^{-2}}; \quad \frac{1%
}{M_{0}} \leq H \leq H_{0} \tag{6.5}
\]
On the basis of (6.2), (6.3), (6.5) we now introduce the following functions
spaces, for $t_{0}\in [0, T[$ and $\delta >0$:
\begin{align*}
&E_{t_{0}}^{\delta }=\left\{ s\in C(t_{0},\,t_{0}+\delta ); \frac{1}{M_{0}}
\leq s \leq \frac{1}{1+2M_{0}^{-2}};\,\frac{1}{\alpha \left(
s(t_{0}+t)\right) }\leq M_{0};t\in [0,\delta ]\right\} \\
&F_{t_{0}}^{\delta }=\left\{ H \in C(t_{0},\,t_{0}+\delta );\frac{1}{M_{0}}%
\leq H(t_{0}+t)\leq H_{o};\,t\in [0,\delta ]\right\}  \tag{6.6} \\
&G_{t_{0}}^{\delta }=\left\{ \Sigma _{+}\in C(t_{0},\,t_{0}+\delta ); -1
\leq \Sigma _{+}(t_{0}+t)\leq \frac{1}{2} ;\,\,\,\,t\in [0,\delta ]\right\}
\end{align*}
One verifies easily that $E_{t_{0}}^{\delta }$, $F_{t_{0}}^{\delta }$, $%
G_{t_{0}}^{\delta }$ are complete metric subspaces of the Banach space
denoted $C(t_{0},\,t_{0}+\delta )$, of the continuous (and hence bounded)
functions on the line segment $[t_{0},\,t_{0}+\delta ]$, endowed with the
norm:
\[
\left\| u\right\| _{\infty }=\sup_{t\in [t_{0},t_{0}+\delta ]}\left|
u(t)\right|; \quad u \in C(t_{0},\,t_{0}+\delta ) \tag{6.7}
\]

\subsection{The global existence theorem}

\begin{proposition}
\label{p:6.2} Let $t_{0}\in [0,T[$.\newline
There exists a strictly positive number $\delta>0$, depending only on the
absolute constants $a_{0},b_{0},\dot{a}_{0},\dot{b}_{0}, T$ and $r$, such
that, the initial value problem for the system (V), with the initial data $(%
\tilde{f}(t_{0}), \tilde{H}(t_{0}), \tilde{s}(t_{0}),
\tilde{z}(t_{0}), \tilde{\Sigma}_{+}(t_{0}))$ at $t=t_{0}$, has a
unique solution $(f,H,s,z,\Sigma_{+}) \in
C([t_{0},\,t_{0}+\delta]; X_{r}) \times F_{t_{0}}^{\delta } \times
E_{t_{0}}^{\delta } \times E_{t_{0}}^{\delta } \times
G_{t_{0}}^{\delta }$.
\end{proposition}
\textbf{Proof:} By theorem \ref{t:3.2}, we know that, if we fix $\bar{s}$, $%
\bar{z} \in E_{t_{0}}^{\delta }$ and if we define $\bar{a}=a(\bar{s},\bar{z}%
) $, $\bar{b}=b(\bar{s},\bar{z})$ by (4.28), then equation (5.1) in $f$ has
a unique solution $f \in C([t_{0},\,t_{0}+\delta]; X_{r})$ such that $%
f(t_{0})=\tilde{f}(t_{0})$, and, by (3.19) and (5.17) with
$f=\tilde{f}$, that:
\[
\|f(t)\| \leq \|\tilde{f}(t_{0})\| \leq \|f_{0}\|< r, \quad t\in [t_{0},
t_{0}+\delta] \tag{6.8}
\]
Next, by proposition \ref{p:4.12}, we know that, there exists a number $%
\delta>0$, (we can suppose $0<\delta<1$), such that, if $\bar{f}$ is given
in $C([t_{0},\,t_{0}+\delta]; X_{r})$, then the system (IV.2) has a unique
solution $(H,s,z,\Sigma_{+})$ on $[t_{0},\,t_{o}+\delta]$ such that $%
(H,s,z,\Sigma_{+})(t_{0})=(\tilde{H},\tilde{s},\tilde{z},\tilde{\Sigma}%
_{+})(t_{0})$. Now proposition \ref{p:6.1} and inequalities (4.51)
show that $(H,s,z,\Sigma_{+}) \in \times F_{t_{0}}^{\delta }
\times E_{t_{0}}^{\delta } \times E_{t_{0}}^{\delta } \times
G_{t_{0}}^{\delta }$. This allows us, setting\\ $X_{t_{0}}^{\delta
}=C([t_{0},\,t_{0}+\delta]; X_{r})\times \left(
E_{t_{0}}^{\delta } \times E_{t_{0}}^{\delta }\right)$ and \\$%
Y_{t_{0}}^{\delta }=C([t_{0},\,t_{0}+\delta] ;X_{r})\times \left(
F_{t_{0}}^{\delta }\times E_{t_{0}}^{\delta }\times E_{t_{0}}^{\delta
}\times G_{t_{0}}^{\delta }\right)$, to define the application:
\begin{equation}
\emph{g}:X_{t_{o}}^{\delta } \rightarrow Y_{t_{o}}^{\delta
};\,\,\,\,\,\,\left( \overline{f},(\overline{s},\overline{z})\right) \mapsto
\left( f,\,(H,\,s,\,z,\,\Sigma _{+})\right)  \tag{6.9}
\end{equation}
We are going to show that, we can find $\delta>0$, such that $\emph{g}$
defined by (6.9) induces a contracting map of the complete metric space $%
X_{t_{0}}^{\delta }$ (defined above) into itself that will hence have a
unique fixed point $(f,s,z)$. This will allow us to find $H$ and $\Sigma_{+}$
such that $(f,(H,s,z,\Sigma_{+}))$ be the unique solution in $%
Y_{t_{0}}^{\delta }$ (defined above) of the initial value problem
for the system (V) with the initial data (6.1).

So, if we evaluate the r.h.s of (5.1) for, $s=\bar{s}$, $z=\bar{z}
\in E_{t_{0}}^{\delta
} $, and if we set, in (5.3) and (5.5) $P_{i}=\bar{P}_{i}=P_{i}(s,z,\bar{f})$%
, i=1, 2; where $\bar{f} \in C([t_{0},\,t_{0}+\delta]; X_{r})$, then there
exists a solution $(f,(H,s,z,\Sigma_{+}))$ of that system, taking the
initial data (6.1) at $t=t_{0}$, or equivalently, a solution of the
following integral system:

\[
\,\,\text{(VI)}\,\,\,\left\{
\begin{array}{l}
f(t_{o}+t)=\widetilde{f}(t_{o})+\int_{t_{o}}^{t_{o}+\,t}\frac{1}{p^{o}(%
\overline{s},\overline{z})}Q(f,\,f)(\overline{s},\,\overline{z})(\tau )d\tau
\\
H(t_{o}+t)=\widetilde{H}(t_{o})+\int_{t_{o}}^{t_{o}+\,t}\left[-\frac{3}{2}%
(1+\Sigma _{+}^{2})H^{2}-\frac{\overline{P}_{1}+2\overline{P}_{2}}{6}%
\right](\tau )d\tau \\
s(t_{o}+t)=\widetilde{s}(t_{o})+\int_{t_{o}}^{t_{o}+\,t}6s(1-s)\Sigma
_{+}H(\tau )d\tau \\
z(t_{o}+t)=\widetilde{z}(t_{o})+\int_{t_{o}}^{t_{o}+\,t}2z(1-z)(1+\Sigma
_{+}-3s\Sigma_{+})H(\tau )d\tau \\
\Sigma _{+}(t_{o}+t)=\widetilde{\Sigma }_{+}(t_{o})+\int_{t_{o}}^{t_{o}+\,t}%
\left[-\frac{3}{2}(1-\Sigma _{+}^{2})H\Sigma _{+}+\frac{\overline{P}_{1})}{6H%
}(\Sigma _{+}-2)+\frac{\overline{P}_{2}}{3H}(\Sigma _{+}+1)\right](\tau
)d\tau
\end{array}
\,\,\,\,\,\,.\right.
\]

To $(\bar{f}_{i},(\bar{s}_{i},\bar{z}_{i})) \in X_{t_{0}}^{\delta
}$, i=1, 2, corresponds the solution\\
$(f_{i},(H_{i},s_{i},z_{i},\Sigma_{+i})) \in Y_{t_{0}}^{\delta }$,
i=1, 2, of the above integral system (VI). Writing the equations
for i=1, i=2 and subtracting yields, using respectively: (5.10)
with $s_{i}=\bar{s}_{i}$, $z_{i}=\bar{z}_{i}$, $\||f_{i}\|| \leq
r$; (5.11) and (5.12) with $f_{i}=\bar{f}_{i}$, $\||\bar{f}_{i}\||
\leq r$; (4.62), definition (6.6) of $E_{t_{0}}^{\delta }$,
$F_{t_{0}}^{\delta }$, (6.7), $0 \leq t \leq \delta$:
\[
\left\| \left| f_{1}-f_{2}\right| \right\| \leq \delta M_{1}\left[ \left\|
\left| f_{1}-f_{2}\right| \right\| +\left\| \overline{s}_{1}-\overline{s}%
_{2}\right\| _{\infty }+\left\| \overline{z}_{1}-\overline{z}_{2}\right\|
_{\infty }\right] \tag{6.10}
\]
\begin{align*}
\left\| H_{1}-H_{2}\right\| _{\infty } &\leq \delta M_{2}\left\| \left|
\overline{f}_{1}-\overline{f}_{2}\right| \right\| \\
&+\delta M_{2}\left[ \left\| H_{1}-H_{2}\right\| _{\infty }+\left\|
s_{1}-s_{2}\right\| _{\infty }+\left\| z_{1}-z_{2}\right\| _{\infty
}+\left\| \Sigma _{+1}-\Sigma _{+2}\right\| _{\infty }\right]  \tag{6.11}
\end{align*}
\[
\left\| s_{1}-s_{2}\right\| _{\infty }\leq \delta M_{3}\left[ \left\|
H_{1}-H_{2}\right\| _{\infty }+\left\| s_{1}-s_{2}\right\| _{\infty
}+\left\| z_{1}-z_{2}\right\| _{\infty }+\left\| \Sigma _{+1}-\Sigma
_{+2}\right\| _{\infty }\right] \tag{6.12}
\]
\[
\left\| z_{1}-z_{2}\right\| _{\infty }\leq \delta M_{4}\left[ \left\|
H_{1}-H_{2}\right\| _{\infty }+\left\| s_{1}-s_{2}\right\| _{\infty
}+\left\| z_{1}-z_{2}\right\| _{\infty }+\left\| \Sigma _{+1}-\Sigma
_{+2}\right\| _{\infty }\right] \tag{6.13}
\]
\begin{align*}
\left\| \Sigma _{+1}-\Sigma _{+2}\right\| _{\infty } &\leq \delta
M_{5}\left\| \left| \overline{f}_{1}-\overline{f}_{2}\right| \right\| \\
&+\delta M_{5}\left[ \left\| H_{1}-H_{2}\right\| _{\infty }+\left\|
s_{1}-s_{2}\right\| _{\infty }+\left\| z_{1}-z_{2}\right\| _{\infty
}+\left\| \Sigma _{+1}-\Sigma _{+2}\right\| _{\infty }\right] .  \tag{6.14}
\end{align*}
where $M_{1}$, $M_{2}$, $M_{3}$, $M_{4}$, $M_{5}$ are absolute constants,
depending only on the absolute constants $a_{0}$, $b_{0}$, $\dot{a}_{0}$, $%
\dot{b}_{0}$, $T$ and $r$.

We keep (6.10) and add (6.11)-(6.12)-(6.13)-(6.14) to obtain:
\begin{align}
& \left[ \left\| H_{1}-H_{2}\right\| _{\infty }+\left\| s_{1}-s_{2}\right\|
_{\infty }+\left\| z_{1}-z_{2}\right\| _{\infty }+\left\| \Sigma
_{+1}-\Sigma _{+2}\right\| _{\infty }\right]  \nonumber \\
\leq & 4\delta (M_{2}+M_{3}+M_{4}+M_{5})\left\| \left| \overline{f}_{1}-%
\overline{f}_{2}\right| \right\|  \nonumber \\
& + 4\delta (M_{2}+M_{3}+M_{4}+M_{5})\left[ \left\| H_{1}-H_{2}\right\|
_{\infty }+\left\| s_{1}-s_{2}\right\| _{\infty }+\left\|
z_{1}-z_{2}\right\| _{\infty }+\left\| \Sigma _{+1}-\Sigma _{+2}\right\|
_{\infty }\right] .  \tag*{(6.15)}
\end{align}
If we choose $\delta$ such that:
\[
0<\delta< Inf (1, \frac{1}{ 16(M_{1}+M_{2}+M_{3}+M_{4}+M_{5})}) \tag{6.16}
\]
which implies that $\delta M_{1}<\frac{1}{4}$ and $4\delta
(M_{2}+M_{3}+M_{4}+M_{5})<\frac{1}{4}$; then (6.10) and (6.15) give:
\[
\left\{
\begin{array}{ll}
\left\| \left| f_{1}-f_{2}\right| \right\| \leq \frac{1}{3}\left[ \left\|
\overline{s}_{1}-\overline{s}_{2}\right\| _{\infty }+\left\| \overline{z}%
_{1}-\overline{z}_{2}\right\| _{\infty }\right] &  \\
\left[ \left\| H_{1}-H_{2}\right\| _{\infty }+\left\| s_{1}-s_{2}\right\|
_{\infty }+\left\| z_{1}-z_{2}\right\| _{\infty }+\left\| \Sigma
_{+1}-\Sigma _{+2}\right\| _{\infty }\right] \leq \frac{1}{3}\left\| \left|
\overline{f}_{1}-\overline{f}_{2}\right| \right\| . &
\end{array}
\right.
\]
and by addition:
\begin{align*}
& \left\| \left| f_{1}-f_{2}\right| \right\| +\left[ \left\|
H_{1}-H_{2}\right\| _{\infty }+\left\| s_{1}-s_{2}\right\| _{\infty
}+\left\| z_{1}-z_{2}\right\| _{\infty }+\left\| \Sigma _{+1}-\Sigma
_{+2}\right\| _{\infty }\right]  \nonumber \\
\leq & \frac{1}{3} \left[ \left\| \left| \overline{f}_{1}-\overline{f}%
_{2}\right| \right\| + \left\| \overline{s}_{1}-\overline{s}_{2}\right\|
_{\infty }+\left\| \overline{z}_{1}-\overline{z}_{2}\right\| _{\infty
}\right]
\end{align*}
from which we deduce:
\[
\left\| \left| f_{1}-f_{2}\right| \right\| +\left\| s_{1}-s_{2}\right\|
_{\infty }+\left\| z_{1}-z_{2}\right\| _{\infty }\leq \frac{1}{3}\left[
\left\| \left| \overline{f}_{1}-\overline{f}_{2}\right| \right\| +\left\|
\overline{s}_{1}-\overline{s}_{2}\right\| _{\infty }+\left\| \overline{z}%
_{1}-\overline{z}_{2}\right\| _{\infty }\right] \tag{6.17}
\]

(6.17) shows that, the map $(\bar{f}, (\bar{s}, \bar{z})) \, \mapsto \,
(f,(s,z))$ is a contracting map from the complete metric space $%
X_{t_{0}}^{\delta}$ into itself, for every number $\delta$ satisfying
(6.16), which shows that $\delta$ depends only on the absolute constants $%
M_{i}$, i=1, 2, 3, 4, 5. This maps has a unique fixed point $(f,(s,z)) \in
X_{t_{0}}^{\delta}$. Now, to determine $H$ and $\Sigma_{+}$, since $s$ is
known, (5.3) determines the product in $H\Sigma_{+}$ in terms of $s$; then,
substituting this product in (5.4) gives $H$ in terms of $s$ and $z$; once $%
H $ is known, (5.3) gives $\Sigma_{+}$, and we obtain the desired unique
solution $(f,(H,s,z,\Sigma_{+})) \in Y_{t_{0}}^{\delta}$. This ends the
proof of proposition \ref{p:6.2} $\blacksquare$

We can then state:
\begin{theorem}\label{t:6.3} The initial values problem for the spatially homogeneous
Einstein-Boltzmann system on a locally rotationally symmetric Bianchi type I
space-time, has a global solution $(f,a,b)$ on $[0, +\infty[$, for suitable
arbitrarily large initial data at $t=0$.
\end{theorem}

\begin{remark}
\label{r:6.4}

\begin{enumerate}
\item  Nowhere in the proof we had to restrict the size of the initial data,
which can then been taken arbitrarily large.

\item  In \cite{n}, which studies the case $a=b$, the author didn't study
the Einstein evolution equations, which are, as we saw in paragraph 4 the
main problem to solve.

\item  The present work extends the global existence result established in
\cite{w}, in the case $a=b$ and strictly positive cosmological constant $%
\Lambda >0$, to the case $\Lambda =0$.

\item  In the future, we plan to prove the geodesic completness
and to relax the hypotheses on the collision kernel.
\end{enumerate}
\end{remark}

\textbf{\underline{Acknowledgement}}. The authors thank A.D.Rendall and
Satyanad Kichenassamy for helpful comments and suggestions. This work was
supported by the Volkswagen Stiftung, Federal Republic of Germany.


\begin{thebibliography}{99}
\bibitem{u}  A. Lichnerowicz; Theories Relativistes de la Gravitatation et
de l'Electromagnetisme. Masson et Cie, Editeurs, 1955.

\bibitem{t}  J. Ehlers; A survey of General Relativity theory, Astrophysics
and Cosmology, ed. W. Israel, (Reidel, Dordrecht, 1973), p. 1-125.

\bibitem{p}  S. W. Hawking and G. F. R. Ellis. The large scale structure of
space-time, Cambridge Monographs on Mathematical Physics, Cambridge
University Press, 1973.

\bibitem{m}  Bancel, D. and Choquet-Bruhat, Y., Uniqueness and local
existence for the Einstein-Maxwell-Boltzmann system, Comm. Math. Phys. 33,
83-96, (1973).

\bibitem{j}  Di Blasio, G.: Differentiability of Spatially Homogeneous
Solutions of the Boltzmann Equation in the non Maxwellian case. Comm. Maths.
Phys. (1974)

\bibitem{v}  P. Hartmann; Ordinary differential equation,Boston, Basel,
Stuttgart, 1982.

\bibitem{k}  Illner, R. and Shinbrot, M., The Boltzmann equation, global
existence for a rare gas in an infinite vacuum. Comm. Math. Phys. 95,
(1984), 217-226.

\bibitem{i}  Diperna, R. J. and Lions P. L., On the Cauchy problem for
Boltzmann equation: Global solutions and weak stability. Annals of
Mathematics, I. 30 (1989), 321-366.

\bibitem{l}  Glassey, R., T. and Strauss, W., Asymptotic stability of the
relativistic Maxwellian. Publ. Math. RIMS Kyoto, 29, (1992), 301-347.

\bibitem{e}  D. Christodoulou, Bounded variation solutions of the
spherically symmetric Eintein-scalar fields equations. Comm. Pure Appl.
Math. 46, (1993), 1131-1220.

\bibitem{r}  Glassey , R. T., Lectures on the Cauchy problem in transport
theory, Department of Mathematics, Indiana University, Bloomington, In
47405, May (1993).

\bibitem{c}  A. D. Rendall, On the nature of singularities in plane
symmetric scalar field Cosmologies, Gen. Relativity Gravitation 27 (1995)
213-221.

\bibitem{n}  Mucha, P., B., Global existence for the Einstein-Boltzmann
equation in the flat Roberson-Walker space-time, Comm. Math. Phys. 203,
107-118, (1999).

\bibitem{o}  Mucha, P., B., Global existence of solutions of the
Einstein-Boltzmann equation in the spatially homogeneous case, in Evolution
Equations Existence, Regularities and Singularities, Banach Center
Publications, volume 52, Institute of Mathematics, Polish Academy of
Sciences, Warzawa, (2000).

\bibitem{x}  Rendall A.D., Uggla c., Dynamics of spatially homogeneous
locally rotationally symmetric solutions of the Einstein-Vlasov equations,
class. Quantum Grav. 17, (2000) 4697-4713, printed in UK.

\bibitem{q}  Straumann, N. ; On the Cosmological constant problems and the
astronomical evidence for homogeneous energy density with negative pressure.
Preprint astro-ph / 0203330, (2002).

\bibitem{a}  H. Andr\'easson: The Einstein-Vlasov system / kinetic theory,
Living Rev. Relativ. 5, lrs-2002-7.

\bibitem{h}  Tchapnda, S. B. and Rendall, A. D., Global existence and
asymptotic behaviour in the future for the Einstein-Vlasov system wih
positive cosmological constant. Class. Quantum Grav. 20, 3037-3049, (2003).

\bibitem{g}  Tchapnda, S. B. and Noutchegueme, N. , the surface symmetric
Einstein-Vlasov system with cosmological constant. Math. Proc.
Camb. Phil. Soc. (2005) 138, 541-553.

\bibitem{f}  Noundjeu, P. and Noutchegueme, N., local existence and
continuation criterion for solutions of the spherically symmetric
Einstein-Vlasov-Maxwell system. General Relativity and
Gravitation, Vol. 36, $N^{o}$6, 1373-1398, (2004).

\bibitem{b}  Rendall, A. D., The Einstein-Vlasov System in the Einstein
Equations and the large Scale Behaviour of the Gravitational Fields, eds. P.
T. Chrusciel and H. Frriedrichs (Birkh$\ddot{a}$user, Basel, 2004).

\bibitem{d}  D. Tegankong, Noutchegueme, N. and Rendall, A. D., Local
Existence and Continuation Criteria for Solutions of the
Einstein-Vlasov-Scalar Field System, JHDE, Voll. 1, N${{}^\circ}$4, 691-724,
(2004).

\bibitem{s}  Noutchegueme, N.; Dongo, D. and Takou, E., Global Existence of
Solutions for the Relativistic Boltzmann Equation with Arbitrarily
Large Initial Data on a Bianchi Type I Space-time. Preprint gr-qc
/ 0503048. To appear in GRG.

\bibitem{w}  Noutchegueme, N; Takou, E; Global existence of solutions for
the Einstein-Boltzmann system with Cosmological constant in a
Robertson-Walker space-time Preprint: gr-qc 0507042.
\end{thebibliography}
\end{document}